\documentclass[prb,preprint]{revtex4}
\usepackage{amssymb}
\usepackage{amsmath}
\usepackage{epsfig}
\usepackage{bbm}
\usepackage{graphicx}

\def\bG{{\bf G}}
\def\bi{{\bf i}}
\def\bj{{\bf j}}
\def\bk{{\bf k}}
\def\bL{{\bf L}}

\def\bq{{\bf q}}

\def\bG{{\bf G}}
\def\bQ{{\bf Q}}
\def\bS{{\bf S}}

\def\bU{{\bf U}}

\def\bV{{\bf V}}

\def\b0{{\bf 0}}

\def\cO{{\cal O}}
\def\cR{{\cal R}}
\def\cS{{\cal S}}
\def\cU{{\cal U}}

\def\Re{{\rm Re}}
\def\Im{{\rm Im}}
\def\bra{\langle}
\def\ket{\rangle}
\def\up{\uparrow}
\def\down{\downarrow}

\def\alf{\alpha}
\def\eps{\epsilon}

\def\Gam{\Gamma}

\def\Lam{\Lambda}

\def\sg{\sigma}
\def\Sg{\Sigma}
\def\bGam{{\bf\Gam}}
\def\bSg{{\bf\Sg}}

\def\phib{\bar\phi}
\def\psib{\bar\psi}

\def\sgn{{\rm sgn}}
\def\det{{\rm det}}

\def\q2{{\textstyle\frac{q}{2}}}

\begin{document}

\title{Effective interactions and fluctuation effects in spin-singlet 
 superfluids}
\date{May 27, 2013}
\author{Andreas Eberlein}
\author{Walter Metzner}
\affiliation{Max Planck Institute for Solid State Research, 
 D-70569 Stuttgart, Germany}

\begin{abstract}
We derive and evaluate one-loop functional flow equations for 
the effective interactions, self-energy and gap function in 
spin-singlet superfluids.
The flow is generated by a fermionic frequency cutoff, which is 
supplemented by an external pairing field to treat divergencies 
associated with the Goldstone boson.
To parametrize the singular momentum and frequency dependences
of the effective interactions, the Nambu interaction vertex is 
decomposed in charge, magnetic, and normal and anomalous pairing 
channels.
The one-loop flow solves reduced (mean-field) models for 
superfluidity exactly, and captures also important fluctuation
effects.
The Ward identity from charge conservation is generally violated,
but can be enforced by projecting the flow.
Applying the general formalism to the two-dimensional attractive
Hubbard model, we obtain detailed results on the momentum and 
frequency dependences of the effective interactions for weak
and moderate bare interactions.
The gap is reduced by fluctuations, with a stronger reduction
at weaker interactions, as expected.

\vskip 5mm

\end{abstract}

\maketitle

\section{Introduction}

Numerous interacting Fermi systems undergo a phase transition
associated with spontaneous symmetry breaking at sufficiently low
temperatures. Mean-field theory captures salient features of the 
symmetry-broken phase such as long-range order, quasi-particle
excitations, and collective modes.
For example, the BCS wave function provides a surprisingly 
faithful qualitative description of the superfluid ground state 
of an attractively interacting Fermi gas not only at weak, but 
also at strong coupling.\cite{eagles69,leggett80}
Nevertheless, fluctuations often play an important role, both
above and below the energy scale for symmetry breaking.
At high energies, they renormalize the effective interactions
generating an instability of the normal (symmetric) state, which 
may enhance or reduce the scale for symmetry breaking.
At low energies, order parameter fluctuations usually suppress 
the order at least partially.
Triggered by the possibility of designing tunable attractively 
interacting Fermi systems in cold atom traps, the issue of
fluctuation effects in fermionic superfluids has attracted 
renewed interest.\cite{zwerger12}

A framework to deal with fluctuation effects on all energy 
scales is provided by the functional renormalization group (fRG). 
This method provides a flexible source of new approximation 
schemes for interacting Fermi systems,\cite{metzner12}
which are obtained by truncating the exact functional flow for the 
effective action as a function of a decreasing infrared cutoff $\Lam$.
\cite{wetterich93,berges02,kopietz10}
The common types of spontaneous symmetry breaking such as 
superconductivity or magnetic order are associated with a divergence 
of the effective two-particle interaction at a finite scale $\Lambda_c$ 
in a specific momentum channel.\cite{zanchi00,halboth00,honerkamp01} 
To continue the flow below the scale $\Lambda_c$, an order parameter
describing the broken symmetry has to be introduced.

A natural procedure is to decouple the interaction by a bosonic 
order parameter field, via a Hubbard-Stratonovich transformation, 
and to study the coupled flow of the fermionic and order parameter
fields.
Thereby order parameter fluctuations and also their interactions 
can be treated rather easily.
This approach to symmetry breaking in the fRG framework has been 
explored already in several works on antiferromagnetic order 
\cite{baier04,krahl09a} and superconductivity.
\cite{birse05,diehl07,krippa07,strack08,floerchinger08,bartosch09}
The bare microscopic interaction can usually be decoupled by
introducing a single boson field.
However, an effective interaction with only one bosonic field 
corresponds to a strongly simplified representation of the
effective two-fermion interaction.
Systems with competing instabilities corresponding to distinct order 
parameters require the introduction of several bosonic fields.
\cite{krahl09b,friederich10}

Alternatively one may explore purely fermionic flows in the 
symmetry-broken phase. 
This can be done by adding an infinitesimal symmetry breaking 
term to the bare action, which is promoted to a finite order 
parameter below the scale $\Lambda_c$.\cite{salmhofer04}
A simple one-loop truncation of the exact fRG flow equation with
self-energy feedback was shown to yield an {\em exact} description 
of symmetry breaking for mean-field models such as the reduced BCS 
model, although the effective two-particle interactions diverge at
the critical scale $\Lambda_c$.\cite{salmhofer04,gersch05}
A subsequent application to the two-dimensional attractive Hubbard 
model showed that the same truncation, with a rather naive 
parametrization of the effective two-particle vertex, yields 
surprisingly accurate results for the superconducting gap at 
weak coupling.\cite{gersch08}
However, the flow could be carried out down to $\Lam = 0$ only for
a symmetry-breaking pairing field $\Delta_0$ above a certain minimal
value. At that value a spurious divergence of the two-particle 
vertex was found. Fortunately, the minimal $\Delta_0$ was rather 
small, more than two orders of magnitude below the size of the 
gap at the end of the flow, and in this sense close to the ideal 
case of an infinitesimal $\Delta_0$.

In this paper we further develop the fermionic fRG for spin-singlet
superfluids as a prototype for a broken continuous symmetry.
We stay with the one-loop truncation used previously, but we derive
and apply a much more accurate parametrization of the momentum and 
frequency dependence of the flowing two-particle vertex, taking all
singularities in the particle-particle and particle-hole channel 
into account.
We build on recent work on the structure of the Nambu two-particle
vertex in a singlet superfluid,\cite{eberlein10}
where constraints from symmetries (especially spin-rotation
invariance) were derived, and insight into the singularities 
associated with superfluidity was gained by analyzing the exact 
fRG flow of a mean-field model with charge and spin forward scattering 
in addition to the reduced BCS interaction.
Furthermore, a decomposition of the Nambu vertex in distinct interaction 
channels was derived, extending the decomposition formulated
by Husemann and Salmhofer \cite{husemann09} for the normal state,
\cite{karrasch08}
which will now be used to separate regular from singular momentum
and frequency dependences.
With an adequate parametrization of the vertex at hand we can fully
explore the performance of the one-loop truncated fermionic RG for
symmetry-breaking beyond mean-field models.
Explicit results for the effective interactions, the self-energy, and
the gap function will be presented for the two-dimensional Hubbard
model with an attractive interaction as a prototypical case.
The Ward identity relating the gap to the vertex in the phase 
fluctuation channel (Goldstone mode) is not consistent with the 
truncated flow. The deviations are small at weak coupling, but 
they increase with the interaction strength.
This problem can be treated by projecting the flow on the manifold
of effective actions which respect the constraint imposed by the 
Ward identity.
We also analyze to what extent effects of the Goldstone mode on 
other channels are captured by the one-loop truncation.

The article is structured as follows.
In Sec.~II the basic one-loop flow equations for the self-energy 
and the Nambu two-particle vertex are written down.
Symmetry properties of the Nambu vertex following from spin 
rotation invariance and discrete symmetries are reviewed in
Sec.~III.
The channel decomposition for spin-singlet superfluids is 
derived in Sec.~IV, and the general structure of the flow
equations is discussed.
In Sec.~V the random phase approximation is revisited in the
framework of the channel decomposed flow equations.
The general formalism is applied to the attractive Hubbard model
in Sec.~VI, with results for the self-energy, the gap function 
and the effective interactions in all channels.
Merits and shortcomings of the channel decomposed one-loop 
flow equations are summarized in the conclusions, Sec.~VII.

%%%%%%%%%%%%%%%%%%%%%%%%%%%%%%%%%%%%%%%%%%%%%%%%%%%%%%%%%%%%%%%%%%%

\section{Truncated flow equations}

We analyze the superfluid ground state of attractively interacting 
spin-$\frac{1}{2}$ fermions. 
The system is specified by a bare action of the form
\begin{equation} \label{S}
 \cS[\psi,\psib] = 
 - \sum_{k,\sg} [ik_0 - \xi(\bk)] \, \psib_{k\sg} \psi_{k\sg} +
 \cU[\psi,\psib] \; ,
\end{equation}
where $\psib_{k\sg}$ and $\psi_{k\sg}$ are Grassmann variables 
associated with creation and annihilation operators, respectively.
The variable $k = (k_0,\bk)$ contains the Matsubara frequency $k_0$ 
in addition to the momentum $\bk$, and $\sg$ denotes the spin 
orientation.
$\xi(\bk) = \eps(\bk) - \mu$ is the single-particle energy 
relative to the chemical potential, and
$\cU[\psi,\psib]$ describes a spin-rotation invariant two-particle 
interaction
\begin{eqnarray} \label{V}
 \cU[\psi,\psib] &=& 
 \frac{1}{4} \sum_{k_i,\sg_i} \big[ 
 U(k_1,k_2,k_3,k_4) \delta_{\sg_1\sg_4} \delta_{\sg_2\sg_3} -
 U(k_1,k_2,k_4,k_3) \delta_{\sg_1\sg_3} \delta_{\sg_2\sg_4}
 \big] \nonumber \\
 &\times&
 \psib_{k_1\sg_1} \psib_{k_2\sg_2} \psi_{k_3\sg_3} \psi_{k_4\sg_4}
 \; .
\end{eqnarray}
Here and below, all temperature and volume factors are incorporated 
in the summation symbols.

Our analysis is based on a truncation of the exact flow equation 
\cite{wetterich93,metzner12} for the effective action 
$\Gam^{\Lam}[\psi,\psib]$, that is, the generating functional
for one-particle irreducible vertex functions in the presence of 
an infrared cutoff $\Lam$. 
The cutoff is implemented by adding a regulator function to the 
inverse of the bare propagator. 
The effective action $\Gam^{\Lam}[\psi,\psib]$ interpolates between 
the regularized bare action at the initial scale $\Lam_0$ and the 
final effective action $\Gam[\psi,\psib]$ in the limit $\Lam \to 0$.
Spontaneous breaking of the $U(1)$ charge symmetry in the 
superfluid state can be treated by adding a small (ultimately 
infinitesimal) symmetry breaking field 
\begin{equation} \label{deltaS}
 \delta \cS[\psi,\psib] = \sum_k \left[
 \Delta_0(k) \psib_{-k\down} \psib_{k\up} +
 \Delta_0^*(k) \psi_{k\up} \psi_{-k\down} \right]
\end{equation}
to the bare action, which is then promoted to a finite order 
parameter in the course of the flow.\cite{salmhofer04}

Expanding the exact functional flow equation for $\Gam^{\Lam}[\psi,\psib]$ 
in powers of the source fields $\psi$ and $\psib$, one obtains a 
hierarchy of flow equations for the n-particle vertex functions.
\cite{metzner12}
We truncate the hierarchy at the two-particle level, including 
however self-energy corrections generated from contractions of 
three-particle terms.\cite{katanin04}
This truncation was used in all previous fermionic fRG studies of 
symmetry breaking.\cite{salmhofer04,gersch05,gersch08,eberlein10}
It is exact for mean-field models.
Our description of the truncation follows closely the presentation
in Ref.~\onlinecite{eberlein10}. 
However, we use notations as in Ref.~\onlinecite{metzner12}, 
where the regulator function is fully included in the two-point 
vertex $\bGam^{(2)\Lam}$, and the sign convention for $\bGam^{(2)\Lam}$ 
and the propagator $\bG^{\Lam}$ differs from that used in 
Ref.~\onlinecite{eberlein10}.

In a superfluid state it is convenient to use Nambu spinors 
$\phi_{ks}$ and $\phib_{ks}$ defined as
\begin{equation} \label{nambu}
 \phib_{k+} = \psib_{k\up}, \quad 
 \phi_{k+}  = \psi_{k\up}, \quad
 \phib_{k-} = \psi_{-k\down}, \quad 
 \phi_{k-}  = \psib_{-k\down}
\end{equation}
instead of $\psi_{k\sg}$ and $\psib_{k\sg}$ as a basis. 
The effective action as a functional of the Nambu fields,
truncated beyond quartic (two-particle) terms, has the form
\begin{eqnarray} \label{Gamma}
 \Gam^{\Lam}[\phi,\phib] &=& 
 \Gam^{(0)\Lam} -
 \sum_k \sum_{s_1,s_2} \Gam_{s_1s_2}^{(2)\Lam}(k) \,
 \phib_{ks_1} \phi_{ks_2} \nonumber \\
 &+& \frac{1}{4} \sum_{k_1,\dots,k_4} \sum_{s_1,\dots,s_4}
 \Gam_{s_1s_2s_3s_4}^{(4)\Lam}(k_1,k_2,k_3,k_4) \,
 \phib_{k_1s_1} \phib_{k_2s_2} \phi_{k_3s_3} \phi_{k_4s_4} \; ,
\end{eqnarray}
where $\Gam^{(0)\Lam}$ yields the grand canonical potential.
For spin-singlet pairing with unbroken spin-rotation invariance 
only terms with an equal number of $\phi$ and $\phib$ fields 
contribute.
The Nambu vertex $\Gam_{s_1s_2s_3s_4}^{(4)\Lam}(k_1,k_2,k_3,k_4)$
is non-zero only for $k_1+k_2=k_3+k_4$, due to translation
invariance.
The (scale-dependent) Nambu propagator $\bG^{\Lam}$ is related
to $\bGam^{(2)\Lam}$ by $(\bG^{\Lam})^{-1} = \bGam^{(2)\Lam}$,
and can be written as a $2 \times 2$ matrix
\begin{equation} \label{bG}
 \bG^{\Lam}(k) = 
 \left( \begin{array}{cc}
 G_{++}^{\Lam}(k) & G_{+-}^{\Lam}(k) \\
 G_{-+}^{\Lam}(k) & G_{--}^{\Lam}(k)
 \end{array} \right) =
 \left( \begin{array}{cc}
 G^{\Lam}(k) & F^{\Lam}(k) \\
 F^{*\Lam}(k) & -G^{\Lam}(-k)
 \end{array} \right) \; .
\end{equation}
The anomalous propagator $F^{\Lam}(k)$ is a symmetric function 
of $k_0$ and $\bk$.
The Nambu self-energy $\bSg^{\Lam}$ is defined by the Dyson
equation $(\bG^{\Lam})^{-1} = (\bG_0^{\Lam})^{-1} - \bSg^{\Lam}$,
where $\bG_0^{\Lam}$ is the bare regularized propagator 
(in presence of $\Delta_0$).
In the superfluid state it has the form
\begin{equation} \label{bSg}
 \bSg^{\Lam}(k) = 
 \left( \begin{array}{cc}
 \Sg^{\Lam}(k) & \Delta_0(k) - \Delta^{\Lam}(k) \\
 \Delta_0^*(k) - \Delta^{*\Lam}(k) & - \Sg^{\Lam}(-k)
 \end{array} \right) \; ,
\end{equation}
where $\Sg^{\Lam}(k)$ is the normal component of the self-energy
and $\Delta^{\Lam}(k)$ is the (flowing) gap function.

The gap function and the Nambu vertex are related by a Ward 
identity following from global charge conservation (see,
for example, Ref.\ \onlinecite{salmhofer04})
\begin{eqnarray} \label{ward1}
 \Delta^{\Lam}(k) - \Delta_0(k) &=&
 \sum_{k'} \sum_{s,s'} \left[
 \Delta_0(k') G^{\Lam}_{s+}(k') G^{\Lam}_{-s'}(k') -
 \Delta_0^*(k') G^{\Lam}_{s-}(k') G^{\Lam}_{+s'}(k') 
 \right] \nonumber \\
 && \times \, \Gam^{(4)\Lam}_{+s's-}(k,k',k',k) \; .  
\end{eqnarray}
The Ward identity implies that some components of the Nambu 
vertex diverge in case of spontaneous symmetry breaking 
($\Delta^{\Lam}$ finite for $\Delta_0 \to 0$), which is a
manifestation of the massless Goldstone boson.

The flow equation for the Nambu self-energy is given by
\begin{equation} \label{floweq_Sg}
 \frac{d}{d\Lam} \Sg_{s_1s_2}^{\Lam}(k) =
 \sum_{k'} \sum_{s'_1,s'_2} S_{s'_2s'_1}^{\Lam}(k')
 \Gam^{(4)\Lam}_{s_1s'_1s'_2s_2}(k,k',k',k) \; ,
\end{equation}
where
\begin{eqnarray} \label{bS}
 \bS^{\Lam}(k) &=& \left. 
 \frac{d}{d\Lam} \bG^{\Lam}(k) \right|_{\bSg^{\Lam} \, {\rm fixed}}
 \nonumber \\
 &=& \left[1 - \bG_0^{\Lam}(k) \bSg^{\Lam}(k) \right]^{-1}
     \frac{d\bG_0^{\Lam}(k)}{d\Lam}
     \left[1 - \bSg^{\Lam}(k) \bG_0^{\Lam}(k) \right]^{-1}
\end{eqnarray}
is the so-called single-scale propagator.
The truncated flow equation for the Nambu vertex (see Fig.~1) reads
\begin{eqnarray} \label{floweq_Gam4}
 \frac{d}{d\Lam}
 \Gam^{(4)\Lam}_{s_1s_2s_3s_4}(k_1,k_2,k_3,k_4) 
 &=& 
 \Pi^{\rm PH,d}_{s_1s_2s_3s_4}(k_1,k_2,k_3,k_4) - 
 \Pi^{\rm PH,cr}_{s_1s_2s_3s_4}(k_1,k_2,k_3,k_4) \nonumber \\ 
 &-& 
 \frac{1}{2} \Pi^{\rm PP}_{s_1s_2s_3s_4}(k_1,k_2,k_3,k_4) \; ,
\end{eqnarray}
where
\begin{eqnarray} \label{bubbles}
 \Pi^{\rm PH,d}_{s_1s_2s_3s_4}(k_1,k_2,k_3,k_4) &=&
 \sum_{p,q} \sum_{s'_1,\dots,s'_4}
 \frac{d}{d\Lam}
 [G^{\Lam}_{s'_1s'_2}(p) G^{\Lam}_{s'_3s'_4}(q)] \nonumber \\
 &\times& 
 \Gam^{(4)\Lam}_{s_1s'_2s'_3s_4}(k_1,p,q,k_4)
 \Gam^{(4)\Lam}_{s'_4s_2s_3s'_1}(q,k_2,k_3,p) \; , \\
 \Pi^{\rm PH,cr}_{s_1s_2s_3s_4}(k_1,k_2,k_3,k_4) &=&
 \sum_{p,q} \sum_{s'_1,\dots,s'_4}
 \frac{d}{d\Lam}
 [G^{\Lam}_{s'_1s'_2}(p) G^{\Lam}_{s'_3s'_4}(q)] \nonumber \\
 &\times& 
 \Gam^{(4)\Lam}_{s_2s'_2s'_3s_4}(k_2,p,q,k_4)
 \Gam^{(4)\Lam}_{s'_4s_1s_3s'_1}(q,k_1,k_3,p) \; , \\
 \Pi^{\rm PP}_{s_1s_2s_3s_4}(k_1,k_2,k_3,k_4) &=&
 \sum_{p,q} \sum_{s'_1,\dots,s'_4}
 \frac{d}{d\Lam}
 [G^{\Lam}_{s'_1s'_2}(p) G^{\Lam}_{s'_3s'_4}(q)] \nonumber \\
 &\times& 
 \Gam^{(4)\Lam}_{s_1s_2s'_3s'_1}(k_1,k_2,q,p)
 \Gam^{(4)\Lam}_{s'_2s'_4s_3s_4}(p,q,k_3,k_4) \; .
\end{eqnarray}
\begin{figure}[tb]
\centerline{\includegraphics[width=9cm]{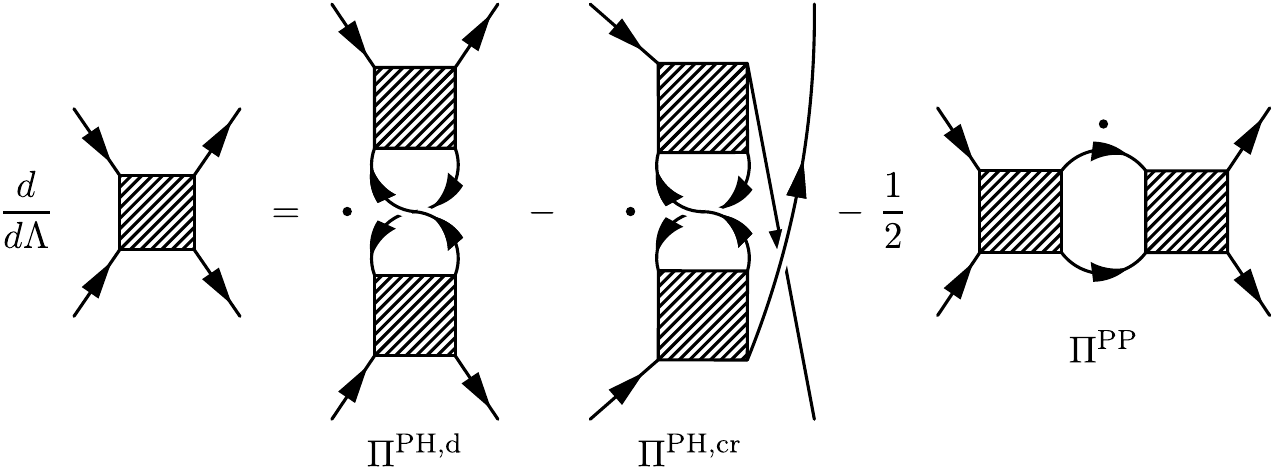}}
\caption{Feynman diagrams representing the flow equation for the
 two-particle vertex. The dots denote differentiation of the 
 propagator products with respect to the scale $\Lam$.}
\end{figure}
The flow equation for the self-energy is exact (for an exact
$\Gam^{(4)\Lam}$), while in the flow of $\Gam^{(4)\Lam}$
contributions from $\Gam^{(6)\Lam}$ beyond self-energy
feedback have been discarded.\cite{katanin04,metzner12}
These discarded contributions are at least of order 
$(\Gam^{(4)\Lam})^3$, and they involve overlapping 
loops leading to a reduced momentum integration volume.
The truncation is exact for mean-field models with a reduced
BCS and/or forward scattering interaction, although
$\Gam^{(4)\Lam}$ becomes large at the critical scale.
\cite{salmhofer04,gersch08,eberlein10}
Particle-particle terms in Nambu representation contain 
particle-hole contributions in the original fermion basis and 
vice versa. In particular, the particle-particle contribution 
generating the Cooper instability is captured by the Nambu 
particle-hole diagrams.

%%%%%%%%%%%%%%%%%%%%%%%%%%%%%%%%%%%%%%%%%%%%%%%%%%%%%%%%%%%%%%%

\section{Symmetries of Nambu vertex}

The Nambu vertex $\Gam_{s_1s_2s_3s_4}^{(4)\Lam}(k_1,k_2,k_3,k_4)$
has 16 components corresponding to the choices $s_i = \pm$ for
$i = 1,\dots,4$.
Spin rotation invariance reduces the number of independent 
components of the Nambu vertex substantially.
In Ref.\ \onlinecite{eberlein10} it was shown that the vertex 
can be parametrized by three functions of $k_1,k_2,k_3,k_4$, 
where $k_1 + k_2 = k_3 + k_4$ for translation invariant systems. 
These functions are further constrained by discrete symmetries.
In this section we describe the spin-rotation invariant form
of the Nambu vertex as derived in Ref.~\onlinecite{eberlein10}.

In addition to the normal interaction, in the $U(1)$ symmetry-broken 
state there are also {\em anomalous} interactions corresponding to 
operator products $\psib\psib\psib\psib$ + conjugate and 
$\psib\psib\psib\psi$ + conjugate.\cite{salmhofer04,gersch08}
Following Ref.~\onlinecite{eberlein10}, we write down 
spin-rotation invariant forms for the normal and anomalous
interaction terms in the $\psi$-basis, and then the corresponding 
expressions in Nambu representation.

A spin-rotation invariant normal interaction can always be 
expressed as \cite{salmhofer01}
\begin{eqnarray} \label{gamma22}
 \Gam^{(2+2)}[\psi,\psib] &=& 
 \frac{1}{4} \sum_{k_i,\sg_i} \left[ 
 V(k_1,k_2,k_3,k_4) \delta_{\sg_1\sg_4} \delta_{\sg_2\sg_3} -
 V(k_1,k_2,k_4,k_3) \delta_{\sg_1\sg_3} \delta_{\sg_2\sg_4}
 \right] \nonumber \\
 &\times&
 \psib_{k_1\sg_1} \psib_{k_2\sg_2} \psi_{k_3\sg_3} \psi_{k_4\sg_4}
 \; .
\end{eqnarray}
Here and in the remainder of this section we suppress the 
superscript $\Lam$ for the scale dependence.
One may also write $\Gam^{(2+2)}[\psi,\psib]$ as a sum of a spin 
singlet and a spin triplet component \cite{halboth00}
\begin{eqnarray}
 \Gam^{(2+2)}[\psi,\psib] &=& 
 \frac{1}{4} \sum_{k_i,\sg_i} \left[ 
 V^S(k_1,k_2,k_3,k_4) \, S_{\sg_1\sg_2\sg_3\sg_4} +
 V^T(k_1,k_2,k_3,k_4) \, T_{\sg_1\sg_2\sg_3\sg_4}
 \right] \nonumber \\
 &\times&
 \psib_{k_1\sg_1} \psib_{k_2\sg_2} \psi_{k_3\sg_3} \psi_{k_4\sg_4}
 \; ,
\end{eqnarray}
where 
$S_{\sg_1\sg_2\sg_3\sg_4} = 
 \frac{1}{2} (\delta_{\sg_1\sg_4} \delta_{\sg_2\sg_3}
 - \delta_{\sg_1\sg_3} \delta_{\sg_2\sg_4})$,
$T_{\sg_1\sg_2\sg_3\sg_4} = 
 \frac{1}{2} (\delta_{\sg_1\sg_4} \delta_{\sg_2\sg_3}
 + \delta_{\sg_1\sg_3} \delta_{\sg_2\sg_4})$, and
\begin{eqnarray}
 V^S(k_1,k_2,k_3,k_4) &=& 
 V(k_1,k_2,k_3,k_4) + V(k_1,k_2,k_4,k_3) \; , \\
 V^T(k_1,k_2,k_3,k_4) &=& 
 V(k_1,k_2,k_3,k_4) - V(k_1,k_2,k_4,k_3) \; .
\end{eqnarray}

A spin-rotation invariant anomalous interaction with four
creation (or annihilation) operators can be written in the 
form \cite{eberlein10}
\begin{eqnarray} \label{gamma40}
 \Gam^{(4+0)}[\psi,\psib] =&& \nonumber \\
 \frac{1}{8} \sum_{k_i} &&
 \Big\{ W^S(k_1,k_2,k_3,k_4) 
 (\psib_{k_1\up}\psib_{k_2\down} - \psib_{k_1\down}\psib_{k_2\up})
 (\psib_{k_3\up}\psib_{k_4\down} - \psib_{k_3\down}\psib_{k_4\up})
 \nonumber \\
  - && \,
 W^T(k_1,k_2,k_3,k_4) \big[
 (\psib_{k_1\up}\psib_{k_2\down} + \psib_{k_1\down}\psib_{k_2\up})
 (\psib_{k_3\up}\psib_{k_4\down} + \psib_{k_3\down}\psib_{k_4\up})
 \nonumber \\[2mm]
 && -2 
 (\psib_{k_1\up} \psib_{k_2\up} \psib_{k_3\down} \psib_{k_4\down} +
  \psib_{k_1\down} \psib_{k_2\down} \psib_{k_3\up} \psib_{k_4\up})
 \big] + {\rm conj.} \Big\} \; .
\end{eqnarray}
Conjugated terms denoted by ''conj.'' are obtained by reversing
the order of fields, replacing $\psib_{k\sg}$ by $\psi_{k^*\sg}$, and
complex conjugation of the functions $W^{S,T}$.

Finally, spin-rotation invariant anomalous interactions with
three creation and one annihilation operators, or vice versa,
can be written as \cite{eberlein10}
\begin{eqnarray} \label{gamma31}
 \Gam^{(3+1)}[\psi,\psib] &=& \frac{1}{2} \sum_{k_i}
 \Big\{ X^S(k_1,k_2,k_3,k_4) 
 \sum_{\sg} 
 \psib_{k_1\sg} (\psib_{k_2\up} \psib_{k_3\down} - 
 \psib_{k_2\down} \psib_{k_3\up}) \psi_{k_4\sg} \nonumber \\
 && + \; X^T(k_1,k_2,k_3,k_4)  
 \Big[ \sum_{\sg} \eps_{\sg}
 \psib_{k_1\sg} (\psib_{k_2\up} \psib_{k_3\down} + 
 \psib_{k_2\down} \psib_{k_3\up}) \psi_{k_4\sg} \nonumber \\
 && \quad + \; 
 2(\psib_{k_1\up}\psib_{k_2\down}\psib_{k_3\down}\psi_{k_4\down}
 - \psib_{k_1\down}\psib_{k_2\up}\psib_{k_3\up}\psi_{k_4\up})
 \Big] + {\rm conj.} \Big\} \; ,
\end{eqnarray}
where $\eps_{\up} = 1$ and $\eps_{\down} = -1$.

It is convenient to collect the 16 components of the Nambu vertex
$\Gam_{s_1s_2s_3s_4}^{(4)}$ in a $4 \times 4$ matrix
\begin{equation} \label{vertexmatrixdef}
 \bGam^{(4)} = \left( \begin{array}{cccc}
 \Gam_{++++}^{(4)} & \Gam_{++-+}^{(4)} & 
 \Gam_{+-++}^{(4)} & \Gam_{+--+}^{(4)} \\[2mm]
 \Gam_{+++-}^{(4)} & \Gam_{++--}^{(4)} & 
 \Gam_{+-+-}^{(4)} & \Gam_{+---}^{(4)} \\[2mm]
 \Gam_{-+++}^{(4)} & \Gam_{-+-+}^{(4)} & 
 \Gam_{--++}^{(4)} & \Gam_{---+}^{(4)} \\[2mm]
 \Gam_{-++-}^{(4)} & \Gam_{-+--}^{(4)} & 
 \Gam_{--+-}^{(4)} & \Gam_{----}^{(4)} 
 \end{array} \right) \quad .
\end{equation}
Rows in this matrix are labeled by $s_1$ and $s_4$, while columns 
are labeled by $s_2$ and $s_3$.
With this convention the Bethe-Salpeter equation yielding the
exact Nambu vertex in reduced (mean-field) models can be written
as a matrix equation.\cite{eberlein10}
Translating the spin-rotation invariant structure of the various 
interaction terms to the Nambu representation,
one obtains the Nambu vertex in the following form \cite{fn2}
\begin{eqnarray} \label{vertexmatrix}
 \bGam^{(4)}(k_1,k_2,k_3,k_4) = \hskip 12.2cm  \nonumber \\[2mm]
 \left( \begin{array}{cccc}
 V^T(k_1,k_2,k_3,k_4) & X(k_1,k_2,k_3,k_4) & 
 X^*(k^*_4,k^*_3,k^*_2,k^*_1) & \!\! -V(k_1,-k_3,-k_2,k_4) \\[2mm]
 \!\! -X(k_1,k_2,k_4,k_3) & W(k_1,k_2,k_3,k_4) & 
 V(k_1,-k_4,-k_2,k_3) & X^*(k_4,k_3,k_1,k_2) \\[2mm]
 \!\! -X^*(k^*_4,k^*_3,k^*_2,k^*_1) & V^*(k_1,-k_4,-k_2,k_3) &
 W^*(k^*_4,k^*_3,k^*_2,k^*_1) & X(k^*_1,k^*_2,k^*_3,k^*_4) \\[2mm]
 \!\! -V^*(k_1,-k_3,-k_2,k_4) & \!\! -X^*(k_4,k_3,k_2,k_1) & 
 \!\! -X(k^*_1,k^*_2,k^*_3,k^*_4) & V^{T*}(k_1,k_2,k_3,k_4) 
 \end{array} \right) \; , \nonumber \\[2mm]
\end{eqnarray}
where $k^* = (-k_0,\bk)$. The matrix elements $W$ and $X$ are related 
to the anomalous (4+0) and (3+1) interactions, respectively:
\begin{eqnarray} \label{W}
 W(k_1,k_2,k_3,k_4) &=& 
 W^S(k_1,-k_4,-k_3,k_2) - W^S(k_1,-k_3,-k_4,k_2)
 \nonumber \\ 
 &+&
 W^T(k_1,-k_4,-k_3,k_2) - W^T(k_1,-k_3,-k_4,k_2)
 \nonumber \\ 
 &+&
 2 W^T(k_1,k_2,-k_3,-k_4) \; ,
\end{eqnarray}
\begin{eqnarray} \label{X}
 X(k_1,k_2,k_3,k_4) &=& 
 X^S(k_1,k_2,-k_3,k_4) - X^S(k_2,k_1,-k_3,k_4)
 \nonumber \\ 
 &+&
 X^T(k_1,k_2,-k_3,k_4) - X^T(k_2,k_1,-k_3,k_4)
 \nonumber \\ 
 &+&
 2 X^T(-k_3,k_2,k_1,k_4) \; .
\end{eqnarray}

For translation invariant systems the functions $V(k_1,k_2,k_3,k_4)$, 
$W(k_1,k_2,k_3,k_4)$ and $X(k_1,k_2,k_3,k_4)$ are non-zero only if 
$k_1 + k_2 = k_3 + k_4$, and can therefore be parametrized by three 
energy and momentum variables.
Discrete symmetries, such as time reversal and reflection invariance,
and the antisymmetry under particle exchange further constrain the
functions parametrizing the Nambu vertex.\cite{eberlein10}

%%%%%%%%%%%%%%%%%%%%%%%%%%%%%%%%%%%%%%%%%%%%%%%%%%%%%%%%%%%%%%%%%%

\section{Channel decomposition}

The two-particle vertex acquires a pronounced momentum and frequency
dependence in the course of the flow, which becomes even singular
at the critical scale for spontaneous symmetry breaking.
A parametrization based on weak coupling power counting is not
adequate in this situation.
Keeping the full dependence on the three independent momenta and
frequencies is technically not feasible.
The particle-particle and particle-hole contributions to the flow, 
Eq.~(\ref{floweq_Gam4}), depend strongly on certain linear 
combinations of momenta and frequencies, namely 
\begin{eqnarray} \label{sing}
 \Pi^{\rm PH,d}_{s_1s_2s_3s_4}(k_1,k_2,k_3,k_4) &:& \; k_3 - k_2 \; ,
 \nonumber \\
 \Pi^{\rm PH,cr}_{s_1s_2s_3s_4}(k_1,k_2,k_3,k_4) &:& \; k_3 - k_1 \; ,
 \nonumber \\
 \Pi^{\rm PP}_{s_1s_2s_3s_4}(k_1,k_2,k_3,k_4) &:& \; k_1 + k_2 \; .
\end{eqnarray}
This is because the poles of the contributing propagators coalesce
when the above combinations of momenta and frequencies vanish or
are situated at special nesting points (in case of nested Fermi 
surfaces).
We therefore write the vertex as a sum of interaction channels,
where each channel carries one potentially singular momentum
dependence, which can be parametrized accurately, while the
dependence on the remaining two momentum variables is treated 
more crudely.
This channel decomposition was introduced by Husemann and Salmhofer 
\cite{husemann09} for the two-particle vertex in a normal metallic
state,\cite{karrasch08} and extended by us for a superfluid state.
\cite{eberlein10}
Most recently it was also formulated for an antiferromagnetic
state.\cite{maier12}

%%%%%%%%%%%%%%%%%%%%%%%%%%%%%%%%%%%%%%%%%%%%%%%%%%%%%%%%%%%%%%%%%%%%%

\subsection{Interaction channels}

Following Husemann and Salmhofer,\cite{husemann09}
we write the normal vertex in the form
\begin{eqnarray} \label{gamma22ch}
 \Gam^{(2+2)\Lam}[\psi,\psib] &=& \cU[\psi,\psib] 
 \nonumber \\ &+& 
 \frac{1}{2} \, {\sum_{k_i}}' \,
 C^{\Lam}_{\frac{k_1+k_4}{2},\frac{k_2+k_3}{2}}(k_3-k_2)
 \sum_{\sg,\sg'}
 \psib_{k_1\sg} \psib_{k_2\sg'} \psi_{k_3\sg'} \psi_{k_4\sg}
 \nonumber \\ &+& 
 \frac{1}{2} \, {\sum_{k_i}}' \,
 M^{\Lam}_{\frac{k_1+k_4}{2},\frac{k_2+k_3}{2}}(k_3-k_2)
 \sum_{\sg_i}
 \vec{\tau}_{\sg_1\sg_4} \cdot \vec{\tau}_{\sg_2\sg_3} \,
 \psib_{k_1\sg_1} \psib_{k_2\sg_2} \psi_{k_3\sg_3} \psi_{k_4\sg_4}
 \nonumber \\ &+& 
 \frac{1}{2} \, {\sum_{k_i}}' \,
 P^{\Lam}_{\frac{k_1-k_2}{2},\frac{k_4-k_3}{2}}(k_1+k_2)
 \sum_{\sg,\sg'}
 \psib_{k_1\sg} \psib_{k_2\sg'} \psi_{k_3\sg'} \psi_{k_4\sg}
 \; , \hskip 5mm
\end{eqnarray}
where $\cU[\psi,\psib]$ is the bare interaction, and the coupling 
functions $C^{\Lam}$, $M^{\Lam}$, and $P^{\Lam}$ capture the 
``charge'', ``magnetic'' (spin), and ``pairing'' channels, 
respectively. The matrices collected in 
$\vec{\tau} = (\tau^{x},\tau^{y}, \tau^{z})$ 
are the three Pauli matrices.
The prime at the sums over $k_i$ indicates momentum (and frequency) 
conservation, $k_1 + k_2 = k_3 + k_4$.
The momentum argument in brackets is the momentum transfer for the
charge and magnetic channels, and the total momentum for the pairing
channel. These are the variables for which a singular dependence is
expected.
Comparing the ansatz Eq.~(\ref{gamma22ch}) to the general 
spin-rotation invariant form of the normal vertex Eq.~(\ref{gamma22}), 
written in terms of $V^{\Lam}(k_1,k_2,k_3,k_4)$, one obtains the 
relation
\begin{eqnarray}
 V^{\Lam}(k_1,k_2,k_3,k_4) &=& U(k_1,k_2,k_3,k_4)
 \nonumber \\ &+& \Big[
 C^{\Lam}_{\frac{k_1+k_4}{2},\frac{k_2+k_3}{2}}(k_3-k_2) + 
 P^{\Lam}_{\frac{k_1-k_2}{2},\frac{k_4-k_3}{2}}(k_1+k_2) 
 \nonumber \\ &-& 
 M^{\Lam}_{\frac{k_1+k_4}{2},\frac{k_2+k_3}{2}}(k_3-k_2) -
 2 M^{\Lam}_{\frac{k_1+k_3}{2},\frac{k_2+k_4}{2}}(k_1-k_3)
 \Big] \delta_{k_1+k_2,k_3+k_4}
 \; .
\end{eqnarray}

The flow equations for $C^{\Lam}$, $M^{\Lam}$ and $P^{\Lam}$
are obtained by choosing a Nambu component involving the normal
interaction $V^{\Lam}$, such as $\Gam_{+-+-}^{(4)\Lam}(k_1,k_2,k_3,k_4) 
= V^{\Lam}(k_1,-k_4,-k_2,k_3)$, and linking the flow of the various
components to the Nambu particle-particle and particle-hole terms
such that momenta in brackets correspond to the strong momentum
dependences as in Eq.~(\ref{sing}).
One thus obtains \cite{eberlein10}
 \begin{eqnarray}
 \label{floweqC}
 \frac{d}{d\Lam} C^{\Lam}_{kk'}(q) &=&
 \frac{1}{4} \Pi^{\rm PP}_{+-+-}(k+\q2,\q2-k,k'+\q2,\q2-k')
 \nonumber \\ &&  - \,
 \Pi^{\rm PH,cr}_{+-+-}(k+\q2,-\q2-k',k-\q2,\q2-k') \; , \\
 \label{floweqM}
 \frac{d}{d\Lam} M^{\Lam}_{kk'}(q) &=&
 \frac{1}{4} \Pi^{\rm PP}_{+-+-}(k+\q2,\q2-k,k'+\q2,\q2-k') \; , \\
 \label{floweqP}
 \frac{d}{d\Lam} P^{\Lam}_{kk'}(q) &=&
 \Pi^{\rm PH,d}_{+-+-}(k+\q2,k'-\q2,k'+\q2,k-\q2) \; .
 \end{eqnarray}
The pairing interaction can be split into a singlet and a triplet
component as
\begin{equation}
 P^{\Lam}_{kk'}(q) = 
 P^{S,\Lam}_{kk'}(q) + P^{T,\Lam}_{kk'}(q) \; ,
\end{equation}
where $P^{S,\Lam}_{kk'}(q)$ is symmetric under sign changes of $k$ 
and $k'$, while $P^{T,\Lam}_{kk'}(q)$ is antisymmetric.

For the anomalous (4+0)-interactions the dependence on the (total)
momentum of the Cooper pairs contained in $\Gam^{(4+0)\Lam}[\psi,\psib]$, 
Eq.~(\ref{gamma40}), is expected to become singular, which is taken
into account by the ansatz
\begin{equation}
 W^{\nu,\Lam}(k_1,k_2,k_3,k_4) =
 W^{\nu,\Lam}_{\frac{k_1-k_2}{2},\frac{k_4-k_3}{2}}(k_1 + k_2) \,
 \delta_{k_1+k_2+k_3+k_4,0}
\end{equation}
for $\nu = S,T$. Eq.~(\ref{W}) then yields
\begin{eqnarray}
 W^{\Lam}(k_1,k_2,k_3,k_4) &=& \Big[
 W^{S,\Lam}_{\frac{k_1+k_4}{2},\frac{k_2+k_3}{2}}(k_3 - k_2) -
 W^{S,\Lam}_{\frac{k_1+k_3}{2},\frac{k_2+k_4}{2}}(k_1 - k_3)
 \nonumber \\ &+& 
 W^{T,\Lam}_{\frac{k_1+k_4}{2},\frac{k_2+k_3}{2}}(k_3 - k_2) -
 W^{T,\Lam}_{\frac{k_1+k_3}{2},\frac{k_2+k_4}{2}}(k_1 - k_3)
 \nonumber \\ &+& 
 2 W^{T,\Lam}_{\frac{k_1-k_2}{2},\frac{k_3-k_4}{2}}(k_1 + k_2)
 \Big] \delta_{k_1+k_2,k_3+k_4} \; .
\end{eqnarray}
A Nambu vertex component capturing this interaction is
$\Gam_{++--}^{(4)\Lam}(k_1,k_2,k_3,k_4)$.
Matching again the strong momentum dependences in brackets with
those of the particle-particle and particle-hole terms, one gets
\cite{eberlein10}
\begin{eqnarray}
 \label{floweqWS}
 \frac{d}{d\Lam} W^{S,\Lam}_{kk'}(q) &=&
 \Pi^{\rm PH,d}_{++--}(k+\q2,k'-\q2,k'+\q2,k-\q2)
 \nonumber \\ &&  - \frac{1}{4}
 \Pi^{\rm PP}_{++--}(k+\q2,\q2-k,\q2-k',k'+\q2) \; , \\
 \label{floweqWT}
 \frac{d}{d\Lam} W^{T,\Lam}_{kk'}(q) &=&
 \frac{1}{4} \Pi^{\rm PP}_{++--}(k+\q2,\q2-k,\q2-k',k'+\q2) \; .
\end{eqnarray}

The functions $X^{S,\Lam}(k_1,k_2,k_3,k_4)$ and 
$X^{T,\Lam}(k_1,k_2,k_3,k_4)$ parametrizing $\Gam^{(3+1)\Lam}[\psi,\psib]$ 
in Eq.~(\ref{gamma31}) are expected to depend singularly on 
$k_2 + k_3$, which is the total momentum of the Cooper pair in 
$\Gam^{(3+1)\Lam}[\psi,\psib]$. We therefore write
\begin{equation}
 X^{\nu,\Lam}(k_1,k_2,k_3,k_4) =
 X^{\nu,\Lam}_{\frac{k_1+k_4}{2},\frac{k_2-k_3}{2}}(k_2 + k_3) \,
 \delta_{k_1+k_2+k_3,k_4}
\end{equation}
for $\nu = S,T$. Eq.~(\ref{X}) then yields
\begin{eqnarray}
 X^{\Lam}(k_1,k_2,k_3,k_4) &=& \Big[
 X^{S,\Lam}_{\frac{k_1+k_4}{2},\frac{k_2+k_3}{2}}(k_2 - k_3) -
 X^{S,\Lam}_{\frac{k_2+k_4}{2},\frac{k_1+k_3}{2}}(k_1 - k_3)
 \nonumber \\ &+& 
 X^{T,\Lam}_{\frac{k_1+k_4}{2},\frac{k_2+k_3}{2}}(k_2 - k_3) -
 X^{T,\Lam}_{\frac{k_2+k_4}{2},\frac{k_1+k_3}{2}}(k_1 - k_3)
 \nonumber \\ &+& 
 2 X^{T,\Lam}_{\frac{k_4-k_3}{2},\frac{k_2-k_1}{2}}(k_1 + k_2)
 \Big] \delta_{k_1+k_2,k_3+k_4} \; .
\end{eqnarray}
Anomalous (3+1)-interactions are contained in the Nambu vertex
component $\Gam_{++-+}^{(4)\Lam}(k_1,k_2,k_3,k_4)$.
Matching singular momentum dependences between the vertex on the 
left hand side and the particle-particle and particle-hole terms 
on the right hand side of the flow equation yields 
\cite{eberlein10}
\begin{eqnarray} 
 \label{floweqXS}
 \frac{d}{d\Lam} X^{S,\Lam}_{kk'}(q) &=&
 \Pi^{\rm PH,d}_{++-+}(k-\q2,k'+\q2,k'-\q2,k+\q2)
 \nonumber \\ &&  - \frac{1}{4}
 \Pi^{\rm PP}_{++-+}(k'+\q2,\q2-k',\q2-k,k+\q2) \; , \\
 \label{floweqXT}
 \frac{d}{d\Lam} X^{T,\Lam}_{kk'}(q) &=&
 \frac{1}{4} \Pi^{\rm PP}_{++-+}(k'+\q2,\q2-k',\q2-k,k+\q2) \; .
\end{eqnarray}

Eqs.~(\ref{floweqC})-(\ref{floweqP}), (\ref{floweqWS}),
(\ref{floweqWT}), (\ref{floweqXS}), and (\ref{floweqXT}) 
determine the flow of the complete set of coupling functions 
describing the Nambu vertex, that is, $C^{\Lam}_{kk'}(q)$, 
$M^{\Lam}_{kk'}(q)$, $P^{\Lam}_{kk'}(q)$, $W^{S,\Lam}_{kk'}(q)$, 
$W^{T,\Lam}_{kk'}(q)$, $X^{S,\Lam}_{kk'}(q)$, and 
$X^{T,\Lam}_{kk'}(q)$, respectively.
Note that the above choice of Nambu components is not unique.
Any component containing $V^{\Lam}$, $W^{\Lam}$, and $X^{\Lam}$,
respectively, could have been chosen. The resulting equations 
for the functions $C^{\Lam}_{kk'}(q)$ etc.\ are equivalent.

Discrete symmetries and Osterwalder-Schrader positivity
(corresponding to hermiticity) constrain the functions 
$C^{\Lam}_{kk'}(q)$ etc.\ by relations analogous to those for 
the interaction functions presented in Sec.~3 of 
Ref.~\onlinecite{eberlein10}.
The normal interaction components obey
\begin{eqnarray}
 C^{\Lam}_{kk'}(q) &=& C^{\Lam}_{\cR k \cR k'}(\cR q) =
 C^{\Lam}_{kk'}(-q) = C^{\Lam *}_{-k,-k'}(q) \; , \\
 M^{\Lam}_{kk'}(q) &=& M^{\Lam}_{\cR k \cR k'}(\cR q) =
 M^{\Lam}_{kk'}(-q) = M^{\Lam *}_{-k,-k'}(q) \; , \\
 P^{\Lam}_{kk'}(q) &=& P^{\Lam}_{\cR k \cR k'}(\cR q) =
 P^{\Lam}_{k'k}(q) = P^{\Lam *}_{-k',-k}(-q) \; ,
\end{eqnarray}
where $\cR k = (k_0,-\bk)$.
Here the first equation follows from inversion symmetry,
the second from inversion and time reversal symmetry, and 
the third from inversion symmetry and positivity.
For the anomalous (4+0)-interaction, invariance under spatial
inversion and time reversal yield the relations
\begin{equation}
 W^{\nu,\Lam}_{kk'}(q) = W^{\nu,\Lam}_{\cR k \cR k'}(\cR q) =
 W^{\nu,\Lam *}_{-k,-k'}(-q)
\end{equation}
for $\nu = S,T$, and for the (3+1)-interactions
\begin{equation}
 X^{\nu,\Lam}_{kk'}(q) = X^{\nu,\Lam}_{\cR k \cR k'}(\cR q) =
 X^{\nu,\Lam *}_{-k,-k'}(-q) \; .
\end{equation}

The complete Nambu vertex can be written in the form (see Fig.~2)
\begin{eqnarray} \label{vertexdecomp}
 \Gam^{(4)\Lam}_{s_1s_2s_3s_4}(k_1,k_2,k_3,k_4) &=&
 U_{s_1s_2s_3s_4}(k_1,k_2,k_3,k_4) \nonumber \\ 
 &+& \Big[ V^{\rm PH,\Lam}_{s_1s_2s_3s_4} \big(
 {\textstyle \frac{k_1+k_4}{2},\frac{k_2+k_3}{2};k_3-k_2} \big)
 - V^{\rm PH,\Lam}_{s_2s_1s_3s_4} \big(
 {\textstyle \frac{k_2+k_4}{2},\frac{k_1+k_3}{2};k_3-k_1} \big)
 \nonumber \\ 
 && + V^{\rm PP,\Lam}_{s_1s_2s_3s_4} \big(
 {\textstyle \frac{k_1-k_2}{2},\frac{k_4-k_3}{2};k_1+k_2} \big)
 \Big] \delta_{k_1+k_2,k_3+k_4} \; ,
\end{eqnarray}
where the first term represents the Nambu components of the
bare interaction, while the other terms are generated by the
particle-hole and particle-particle contributions to the flow,
that is
\begin{eqnarray}
 \frac{d}{d\Lam} V^{\rm PH,\Lam}_{s_1s_2s_3s_4} \big(
 {\textstyle \frac{k_1+k_4}{2},\frac{k_2+k_3}{2};k_3-k_2} \big) 
 &=& \Pi^{\rm PH,d}_{s_1s_2s_3s_4}(k_1,k_2,k_3,k_4) \; ,
 \label{flowVPH} \\
 \frac{d}{d\Lam} V^{\rm PP,\Lam}_{s_1s_2s_3s_4} \big(
 {\textstyle \frac{k_1-k_2}{2},\frac{k_4-k_3}{2};k_1+k_2} \big)
 &=& - \frac{1}{2} 
 \Pi^{\rm PP}_{s_1s_2s_3s_4}(k_1,k_2,k_3,k_4) \; .
 \label{flowVPP}
\end{eqnarray}
The crossed particle-hole contribution yields the flow of 
$V^{\rm PH,\Lam}$ with indices 1 and 2 exchanged and a minus 
sign compared to the direct contribution.
\begin{figure}[tb]
\centerline{\includegraphics[width=9cm]{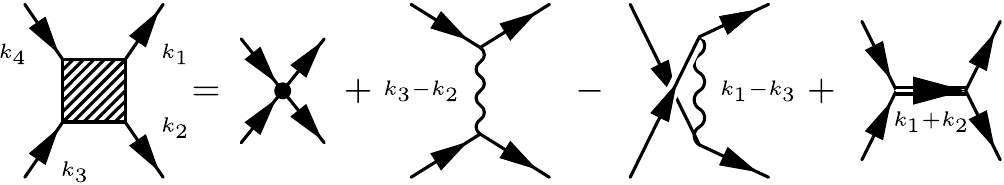}}
\caption{Decomposition of the Nambu vertex in bare interaction,
 particle-hole channels, and particle-particle channel.}
\end{figure}
Collecting terms with the variable $q = k_3-k_2$ in the channel
decomposition, and writing the components in matrix form as in
Eq.~(\ref{vertexmatrixdef}), one obtains
\begin{equation} \label{VPH}
 \bV^{\rm PH,\Lam}(k,k';q) = 
 \left( \begin{array}{cccc}
 K^{+,\Lam}_{kk'}(q) & X^{\Lam}_{kk'}(-q) & 
 X^{\Lam}_{kk'}(q) & -K^{-,\Lam}_{k,-k'}(-q) \\[2mm]
 X^{\Lam}_{k'k}(q) & W^{\Lam}_{kk'}(q) & 
 P^{\Lam}_{kk'}(q) & -X^{\Lam*}_{k'k}(-q) \\[2mm]
 X^{\Lam}_{k'k}(-q) & P^{\Lam*}_{kk'}(q) & 
 W^{\Lam*}_{kk'}(q) &  -X^{\Lam*}_{k'k}(q) \\[2mm]
 -K^{-,\Lam*}_{k,-k'}(-q) & -X^{\Lam*}_{kk'}(q) & 
 -X^{\Lam*}_{kk'}(-q) & K^{+,\Lam*}_{kk'}(q)
 \end{array} \right) \; ,
\end{equation}
where
\begin{eqnarray}
 W^{\Lam}_{kk'}(q) &=& W^{S,\Lam}_{kk'}(q) + W^{T,\Lam}_{kk'}(q) \\
 X^{\Lam}_{kk'}(q) &=& X^{S,\Lam}_{kk'}(q) + X^{T,\Lam}_{kk'}(q) \; ,
\end{eqnarray}
and
\begin{equation}
 K^{\pm,\Lam}_{kk'}(q) = 
 C^{\Lam}_{kk'}(q) \pm M^{\Lam}_{kk'}(q) \; .
\end{equation}
Collecting terms with the variable $q = k_1 + k_2$, one finds
\begin{equation} \label{VPP}
 \bV^{\rm PP,\Lam}(k,k';q) = 
 2 \left( \begin{array}{cccc}
 P^{T,\Lam}_{kk'}(q) & -X^{T,\Lam}_{k'k}(q) & 
 -X^{T,\Lam}_{kk'}(q) & M^{\Lam}_{k,k'}(q) \\[2mm]
 X^{T,\Lam}_{-k',k}(q) & -W^{T,\Lam}_{kk'}(q) & 
 -M^{\Lam}_{k,-k'}(q) & -X^{T,\Lam*}_{-k,k'}(q) \\[2mm]
 X^{T,\Lam}_{-k,k'}(q) & -M^{\Lam}_{k,-k'}(q) & 
 -W^{T,\Lam*}_{kk'}(q) &  -X^{T,\Lam*}_{-k',k}(q) \\[2mm]
 M^{\Lam*}_{kk'}(q) & X^{T,\Lam*}_{kk'}(q) & 
 X^{T,\Lam*}_{k'k}(q) & P^{T,\Lam*}_{kk'}(q)
 \end{array} \right) \; .
\end{equation}
Note that $\bV^{\rm PH,\Lam}$ captures the full information on the
coupling functions $C^{\Lam}_{kk'}(q)$ etc.\ characterizing the Nambu 
vertex. By contrast, $\bV^{\rm PP,\Lam}$ collects only magnetic and 
triplet pairing components.

For $q=0$ the matrix $\bV^{\rm PH,\Lam}$ has the same structure as
the Nambu vertex for a mean-field model with reduced BCS and
forward scattering interactions.\cite{eberlein10}
Contributions with $q \neq 0$ correspond to {\em fluctuations}
away from the zero momentum Cooper and forward scattering
channels.

It is convenient to use linear combinations of $P^{\Lam}$ and 
$W^{\Lam}$ corresponding to amplitude and phase variables.
For a real gap function these combinations are
\begin{eqnarray}
 A^{\Lam}_{kk'}(q) &=& 
 \Re[ P^{\Lam}_{kk'}(q) + W^{\Lam}_{kk'}(q)] \; , \\
 \Phi^{\Lam}_{kk'}(q) &=&
 \Re[ P^{\Lam}_{kk'}(q) - W^{\Lam}_{kk'}(q)] \; .
\end{eqnarray}
Amplitude and phase variables for singlet and triplet components 
can be defined by analogous linear combinations.
Note that $P^{\Lam}_{kk'}(q)$ and $W^{\Lam}_{kk'}(q)$ are 
generally complex functions for $q_0 \neq 0$, even for a real gap.
For their real and imaginary parts we use the notation
$P'^{\Lam}_{kk'}(q)$, $W'^{\Lam}_{kk'}(q)$ and
$P''^{\Lam}_{kk'}(q)$, $W''^{\Lam}_{kk'}(q)$, respectively.
Instead of the representation Eq.~(\ref{vertexmatrixdef}) it can
be advantageous to use a Pauli matrix basis to represent the
Nambu vertex, as described in Appendix A.

%%%%%%%%%%%%%%%%%%%%%%%%%%%%%%%%%%%%%%%%%%%%%%%%%%%%%%%%%%%%%%%%%%%%%%%%%%

\subsection{Boson propagators and fermion-boson vertices}

To achieve an efficient parametrization of the momentum and frequency
dependences, the coupling functions are written in the form of boson 
mediated interactions with bosonic propagators and fermion-boson vertices, 
as proposed by Husemann and Salmhofer.\cite{husemann09}
The bosonic propagators capture the (potentially) singular dependence
on the transfer momentum and frequency while the fermion-boson vertices 
describe the more regular remaining momentum and frequency dependences.
For example, the charge coupling function is decomposed as
\begin{equation} \label{decomp}
 C^{\Lam}_{kk'}(q) = 
 \sum_{\alf,\alf'} C^{\Lam}_{\alf\alf'}(q) \,
 g_{c\alf}^{\Lam}(k,q) g_{c\alf'}^{\Lam}(k',q) \; ,
\end{equation}
where the functions $g_{c\alf}^{\Lam}(k,q)$ provide a real orthonormal
basis set of $k$-space functions, satisfying
\begin{equation} \label{orthonorm}
 \int d\mu(k) \, g_{c\alf}^{\Lam}(k,q) \, g_{c\alf'}^{\Lam}(k,q) 
 = \delta_{\alf\alf'}
\end{equation}
with a suitable (not yet specified) $k$-space measure $d\mu(k)$.
Viewing $C^{\Lam}_{kk'}(q)$ as a boson mediated interaction, the
functions $C^{\Lam}_{\alf\alf'}(q)$ can be interpreted as boson
propagators and $g_{c\alf}^{\Lam}(k,q)$ as fermion-boson vertices.
Analogous decompositions are used for the magnetic and pairing 
coupling functions $M^{\Lam}_{kk'}(q)$ and $P^{\Lam}_{kk'}(q)$,
or the singlet/triplet components of the latter. The anomalous
(4+0) coupling function $W^{\Lam}_{kk'}(q)$ can also be decomposed
in the form Eq.~(\ref{decomp}). Alternatively one may decompose
the amplitude and phase coupling functions.
The anomalous (3+1) coupling functions $X^{\Lam}_{kk'}(q)$ 
require a more general decomposition
\begin{equation} \label{decompX}
 X^{\Lam}_{kk'}(q) = 
 \sum_{\alf,\alf'} X^{\Lam}_{\alf\alf'}(q) \,
 g_{x\alf}^{\Lam}(k,q) \tilde g_{x\alf'}^{\Lam}(k',q) \; ,
\end{equation}
with two different sets of basis functions $g_{x\alf}^{\Lam}$
and $\tilde g_{x\alf}^{\Lam}$, since the $k$ and $k'$-dependence 
of $X^{\Lam}_{kk'}(q)$ is generally different.

Summing over a complete set of basis functions, the above 
decomposition is exact. In practice one has to approximate the
infinite sum by a finite number of terms, with a suitable choice
of boson-propagators and fermion-boson vertices.

%%%%%%%%%%%%%%%%%%%%%%%%%%%%%%%%%%%%%%%%%%%%%%%%%%%%%%%%%%%%%%%%%%%%%

\subsection{Classification of contributions to the flow}

Inserting the channel decomposed Nambu vertex on the right hand side
of the flow equation yields several contributions which can be
distinguished by their topology when representing the coupling
functions by boson mediated interactions.
For a graphical representation we use the symbolic notation from
Fig.~2, where bosons mediating interactions in the (Nambu) 
particle-hole and particle-particle channels are represented by a 
wiggly and a double line, respectively.
All contributions to the flow of the vertex are of second order in
the interaction. We discuss the different topologies for diagrams 
with two wiggly lines as examples.
There are three distinct classes, which we refer to as ``propagator 
renormalization'', ``vertex correction'', and ``box contribution''.
For the propagator renormalization (Fig.~3, left) the momenta
of both bosonic propagators coincide with the momentum transported
through the fermionic bubble. Hence, a singularity in the 
bosonic propagators generated by the bubble is amplified by 
feedback of both propagators. 
For the vertex correction (Fig.~3, right) the momentum of one of 
the bosonic propagators coincides with the momentum of the 
fermionic (Nambu) particle-hole pair.
Potential singularities of the other bosonic propagator are wiped 
out by the momentum integration. 
Note that at zero temperature all expected singularities of 
the vertex are integrable in two spatial dimensions, even the 
infrared singularity associated with the Goldstone mode.
\begin{figure}[tb]
\centerline{\includegraphics[width=3.5cm]{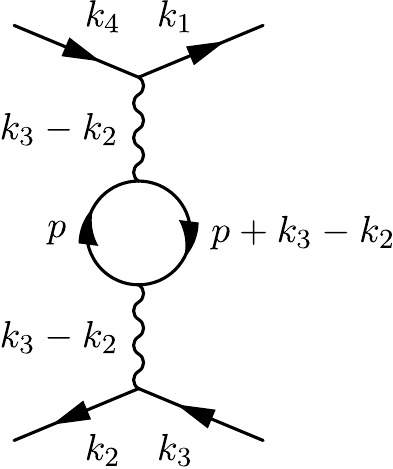}
 \hskip 1cm \includegraphics[width=3cm]{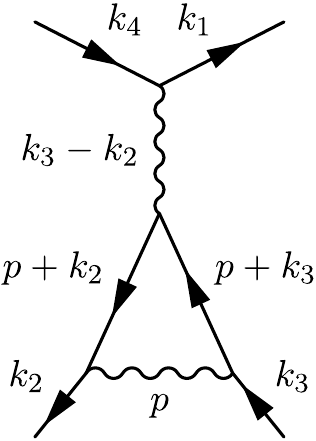}}
\caption{Examples for propagator renormalizaton (left) and vertex
 correction (right). The variable $p$ is integrated.}
\end{figure}
For the box contribution (Fig.~4) singularities of the fermionic 
pair are not amplified by singularities of the bosonic propagators 
which are both integrated.
The contribution from the propagator renormalization diagram
thus dominates in the formation of singularities at special wave 
vectors $q = k_3 - k_2$.
In mean-field models with reduced interactions it yields the
complete flow, while vertex corrections and box contributions
vanish.
\begin{figure}[tb]
\centerline{\includegraphics[width=7cm]{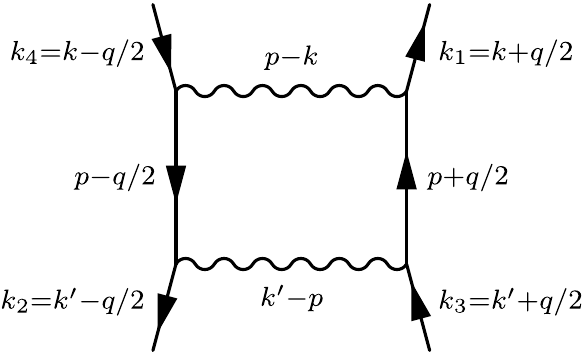}}
\caption{Example for a box diagram with integration variable $p$.}
\end{figure}

The assignment of momenta in the channel decomposition was
designed to deal with singularities generated by the fermionic 
propagators.
However, the box contribution exhibits another singularity
generated by the singularity of the bosonic propagators at
momentum zero. For $k_1 = k_3$ (that is, $k = k'$) the two
bosonic propagators in Fig.~4 carry the same momentum variable.
In the phase fluctuation channel (Goldstone mode) these propagators 
diverge quadratically at small momenta and frequencies (for
$\Delta_0 = 0$).
The product of two such singularities is not integrable in two
dimensions.
This problem can be treated by introducing a scale dependent 
pairing field $\Delta_0^{\Lam}$, which tends to zero continuously
toward the end of the flow. 
A finite pairing field regularizes divergences in the Cooper
channel (including the Goldstone mode), such that the right hand 
side of the flow equations remains finite at each finite scale,
and the flow is integrable down to $\Lam \to 0$, $\Delta_0^{\Lam} \to 0$,
as discussed in more detail in Sec.~\ref{delta0flow}.
A scale dependent $\Delta_0^{\Lam}$ does not modify the structure 
of the flow equations, it merely yields additional contributions to 
the scale derivative of the bare (Nambu) propagator $\bG_0^{\Lam}$.

In addition to the contributions shown in Figs.~3 and 4 there 
are analogous contributions with wiggly lines replaced by double 
lines corresponding to the particle-particle channel and 4-point
vertices representing the bare interaction (as in Fig.~2), 
including all possible mixtures of channels.
The complete set of contributions to the flow of $V^{\rm PH,\Lam}$
and $V^{\rm PP,\Lam}$ is shown in Figs.\ 5 and 6, respectively.
\begin{figure}[tb]
\centerline{\includegraphics[width=8.2cm]{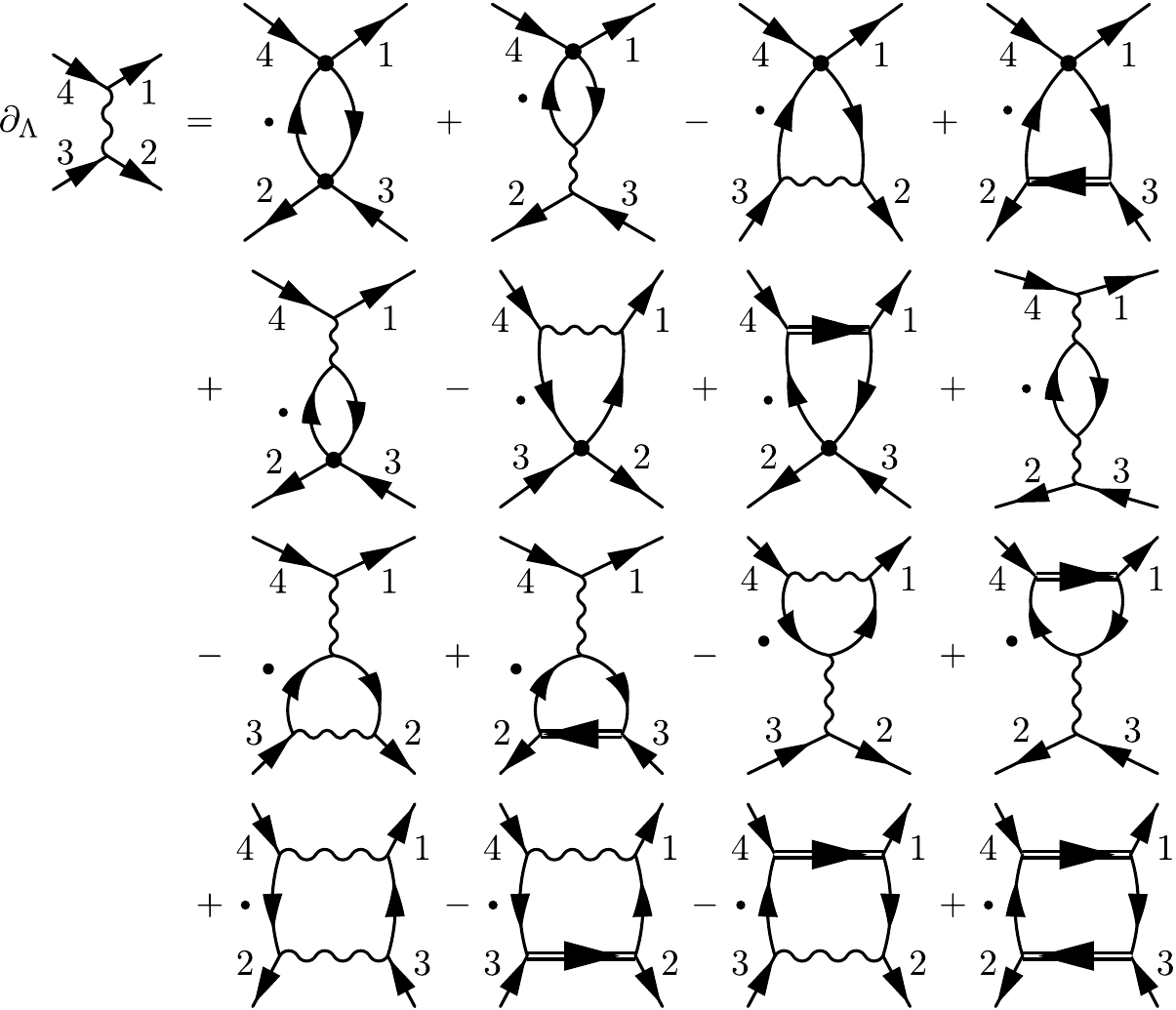}}
\caption{Diagrammatic representation of contributions to the
 flow of $V^{\rm PH,\Lam}$. The dot denotes a $\Lam$-derivative
 acting on the product of the two fermionic propagators.}
\end{figure}
\begin{figure}[tb]
\centerline{\includegraphics[width=8.2cm]{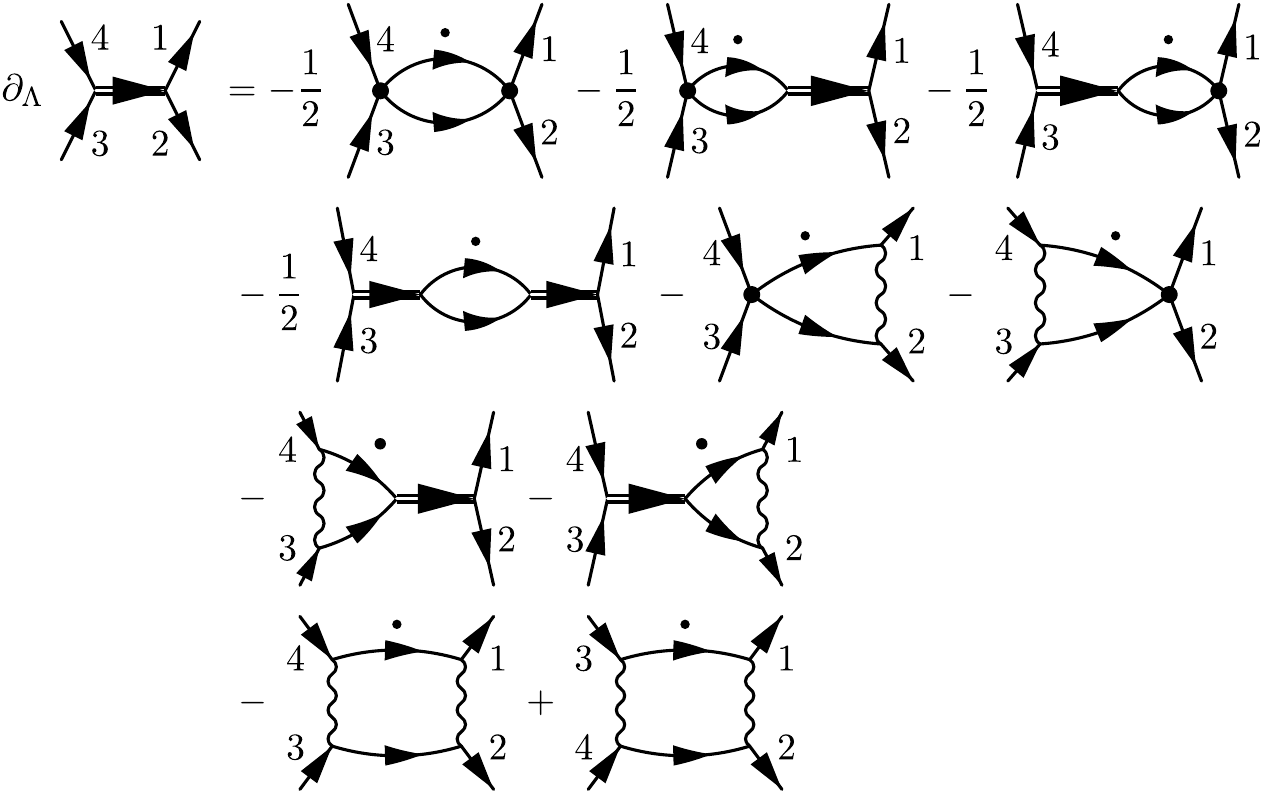}}
\caption{Diagrammatic representation of contributions to the
 flow of $V^{\rm PP,\Lam}$.}
\end{figure}
Note that all diagrams are one-particle irreducible, that is,
they cannot be cut by cutting a single fermionic propagator
line. Some of them can be cut by cutting an interaction line,
but these lines do not correspond to particle propagators, since
the interaction is not represented by bosonic fields in our purely
fermionic RG.

%%%%%%%%%%%%%%%%%%%%%%%%%%%%%%%%%%%%%%%%%%%%%%%%%%%%%%%%%%%%%%%%%%

\section{Random phase approximation}

To gain insight into the singularity structure of the Nambu vertex
it is instructive to consider the random phase approximation (RPA)
before analyzing the full set of flow equations.
In the conventional formulation the RPA corresponds to a summation
of all (direct) particle-hole ladder contributions to the Nambu
vertex with bare interactions and mean-field propagators.
The self-energy is obtained from the usual mean-field equation,
that is, self-consistent first order perturbation theory.
In the channel decomposed functional RG framework derived in 
Sec.~IV, the RPA is equivalent to the approximation
\begin{equation} \label{rpa_vertex}
 \Gam^{(4)\Lam}_{s_1s_2s_3s_4}(k,k';q) = 
 U_{s_1s_2s_3s_4}(k,k';q) + V^{\rm PH,\Lam}_{s_1s_2s_3s_4}(k,k';q)
 \; ,
\end{equation}
that is, the crossed particle-hole and particle-particle channels 
are discarded.
Throughout this section we parametrize the momentum variables
$k_1,k_2,k_3,k_4$ as $k_1 = k + \frac{q}{2}$, 
$k_2 = k' - \frac{q}{2}$, $k_3 = k' + \frac{q}{2}$, and 
$k_4 = k - \frac{q}{2}$.
The flow equation (\ref{flowVPH}) for $V^{\rm PH,\Lam}$ can then
be formally integrated to obtain an integral equation which,
expressed in term of $\Gam^{(4)\Lam}$, reads
\begin{eqnarray} \label{BS}
 \Gam^{(4)\Lam}_{s_1s_2s_3s_4}(k,k';q) &=& 
 U_{s_1s_2s_3s_4}(k,k';q) \nonumber \\ &+&
 \sum_{p} \sum_{s'_i} U_{s_1s'_2s'_3s_4}(k,p;q) \,
 G^{\Lam}_{s'_1s'_2}(p-\q2) 
 G^{\Lam}_{s'_3s'_4}(p+\q2) \,
 \Gam^{(4)\Lam}_{s'_4s_2s_3s'_1}(p,k';q) \; . \hskip 1cm
\end{eqnarray}
This is the familiar Bethe-Salpeter-type equation corresponding
to a summation of (Nambu) particle-hole ladders.
Using this equation, the flow equation for the self-energy 
Eq.~(\ref{floweq_Sg}) can also be integrated, yielding the
usual mean-field equation
\begin{equation}
 \Sg^{\Lam}_{s_1s_2} = \sum_{k'} \sum_{s'_1,s'_2}
 U_{s_1s'_1s'_2s_2}(k,k';0) \, G^{\Lam}_{s'_2s'_1}(k') \; .
\end{equation}
The integral equation (\ref{BS}) can be written in matrix
form such that Nambu index sums correspond to matrix products.
In particular, choosing the Pauli matrix basis described in 
Appendix A, one obtains
\begin{equation} \label{BS_pauli}
 \tilde\bGam^{(4)\Lam}(k,k';q) = \tilde\bU(k,k';q) + 
 \sum_p \tilde\bU(k,p;q) \, \tilde\bL^{\Lam}(p;q) \, 
 \tilde\bGam^{(4)\Lam}(p,k';q) \; ,
\end{equation}
where the components of $\tilde\bL^{\Lam}(p;q)$ are given by
\begin{equation}
 \tilde L^{\Lam}_{jj'}(p;q) = \frac{1}{2} \sum_{s_i}
 \tau^{(j)}_{s_4s_1} \tau^{(j')}_{s_3s_2} \,
 G^{\Lam}_{s_2s_4}(p-\q2) G^{\Lam}_{s_1s_3}(p+\q2) \; .
\end{equation}

For a spin-rotation invariant system, the bare interaction 
can be written in the form
\begin{eqnarray} \label{rpamodel}
 \cU[\psi,\psib] &=&
 \frac{1}{2} \, \sum_{k,k',q} C^{(0)}_{kk'}(q) \sum_{\sg,\sg'}
 \psib_{k+q/2,\sg} \psib_{k'-q/2,\sg'} 
 \psi_{k'+q/2,\sg'} \psi_{k-q/2,\sg}
 \nonumber \\ &+& 
 \frac{1}{2} \, \sum_{k,k',q} M^{(0)}_{kk'}(q) \sum_{\sg_i}
 \vec{\tau}_{\sg_1\sg_4} \cdot \vec{\tau}_{\sg_2\sg_3} \,
 \psib_{k+q/2,\sg_1} \psib_{k'-q/2,\sg_2} 
 \psi_{k'+q/2,\sg_3} \psi_{k-q/2,\sg_4}
 \nonumber \\ &+& 
 \frac{1}{2} \, \sum_{k,k',q} \, P^{(0)}_{kk'}(q) \sum_{\sg,\sg'}
 \psib_{q/2+k,\sg} \psib_{q/2-k,\sg'} 
 \psi_{q/2-k',\sg'} \psi_{q/2+k',\sg}
 \; , \hskip 5mm
\end{eqnarray}
which is analogous to the decomposition of the fluctuation 
contributions in Eq.~(\ref{gamma22ch}).
%Note that the decomposition of the bare interaction in charge,
%magnetic and pairing channels is not unique.
In the special case where the bare coupling functions 
$C^{(0)}_{kk'}(q)$, $M^{(0)}_{kk'}(q)$, and $P^{(0)}_{kk'}(q)$ 
are non-zero only for $q=0$, this becomes the reduced BCS and 
forward scattering interaction of the model discussed in detail 
in Ref.~\onlinecite{eberlein10}. In that case the mean-field
equation for the self-energy is exact, and the Bethe-Salpeter 
equation yields the exact vertex 
$\Gam^{(4)\Lam}_{s_1s_2s_3s_4}(k,k';q=0)$.

For an explicit evaluation of the RPA vertex we assume
separable interactions 
\begin{eqnarray}
 C^{(0)}_{kk'}(q) &=& C^{(0)}(q) f_c(k) f_c(k') \; , \nonumber \\
 M^{(0)}_{kk'}(q) &=& M^{(0)}(q) f_m(k) f_m(k') \; , \nonumber \\
 P^{(0)}_{kk'}(q) &=& P^{(0)}(q) f_p(k) f_p(k') \; ,
\end{eqnarray}
with symmetric (under $k \mapsto -k$) form factors, and a bare
gap function $\Delta_0(k) = \Delta_0 f_p(k)$ with the same form
factor as the pairing interaction.
The coupling functions contributing to $\bV^{\rm PH,\Lam}$, see
Eq.~(\ref{VPH}), then also factorize:
\begin{eqnarray} \label{rpa_fact}
 C^{\Lam}_{kk'}(q) &=& C^{\Lam}(q) f_c(k) f_c(k') \; , \nonumber \\
 M^{\Lam}_{kk'}(q) &=& M^{\Lam}(q) f_m(k) f_m(k') \; , \nonumber \\
 P^{\Lam}_{kk'}(q) &=& P^{\Lam}(q) f_p(k) f_p(k') \; , \nonumber \\
 W^{\Lam}_{kk'}(q) &=& W^{\Lam}(q) f_p(k) f_p(k') \; , \nonumber \\
 X^{\Lam}_{kk'}(q) &=& X^{\Lam}(q) f_c(k) f_p(k') \; ,
\end{eqnarray}
and the gap function has the form
\begin{equation}
 \Delta^{\Lam}(k) = \Delta^{\Lam} f_p(k) \; .
\end{equation}
The vertex assumes a particularly simple form in the Pauli matrix 
basis, namely
\begin{equation}
 \tilde\bGam^{(4)\Lam}(k,k';q) = 
 \tilde{\bf f}(k) \, \tilde\bGam^{(4)\Lam}(q) \, 
 \tilde{\bf f}(k') \; ,
\end{equation}
where $\tilde{\bf f}(k)$ is the diagonal matrix
\begin{equation}
 \tilde{\bf f}(k) = {\rm diag}[f_m(k),f_p(k),f_p(k),f_c(k)] \; ,
\end{equation}
and
\begin{equation}
 \tilde\bGam^{(4)\Lam}(q) = 
 \tilde\bU(q) + \tilde\bV^{\rm PH,\Lam}(q) \; ,
\end{equation}
with
\begin{equation}
 \tilde\bU(q) = 
 \left( \begin{array}{cccc}
 2 M^{(0)}(q) & 0 & 0 & 0 \\
 0 & P^{(0)}(q) & 0 & 0 \\
 0 & 0 & P^{(0)}(q) & 0 \\
 0 & 0 & 0 & 2 C^{(0)}(q) 
 \end{array} \right) 
\end{equation}
and
\begin{equation}
 \tilde\bV^{\rm PH,\Lam}(q) =
 \left( \begin{array}{cccc}
 2 M^{\Lam}(q) & 0 & 0 & 0 \\
 0 & A^{\Lam}(q) & P''^{\Lam}(q) & 2X'^{\Lam}(q) \\
 0 & - P''^{\Lam}(q) & \Phi^{\Lam}(q) &  - 2X''^{\Lam}(q) \\
 0 & 2X'^{\Lam}(q) & 2X''^{\Lam}(q) & 2 C^{\Lam}(q)
 \end{array} \right) \; .
\end{equation}
Here $A^{\Lam}(q) = P'^{\Lam}(q) + W'^{\Lam}(q)$ and
$\Phi^{\Lam}(q) = P'^{\Lam}(q) - W'^{\Lam}(q)$.
Primes denote real parts and double primes imaginary parts.
At $q_0 = 0$ all imaginary parts vanish.
For $q = 0$ the above matrix simplifies to the vertex previously 
obtained for the reduced BCS and forward scattering model,
\cite{eberlein10} in a slightly different basis yielding some
sign changes.

Inserting the factorized form of the vertex into the Bethe-Salpeter
equation Eq.~(\ref{BS_pauli}), one obtains a linear algebraic 
equation for $\tilde\bGam^{(4)\Lam}(q)$,
\begin{equation} \label{rpa_Gamq}
 \tilde\bGam^{(4)\Lam}(q) = \tilde\bU(q) + 
 \tilde\bU(q) \, \tilde\bL^{\Lam}(q) \, \tilde\bGam^{(4)\Lam}(q) \; ,
\end{equation}
where
\begin{equation}
 \tilde\bL^{\Lam}(q) = \sum_p 
 \tilde{\bf f}(p) \tilde\bL^{\Lam}(p;q) \tilde{\bf f}(p) =
 \left( \begin{array}{cccc}
 \tilde L_m^{\Lam}(q) & 0 & 0 & 0 \\
 0 & \tilde L_a^{\Lam}(q) & 
 \tilde L_p''^{\Lam}(q) & \tilde L_x'^{\Lam}(q) \\
 0 & - \tilde L_p''^{\Lam}(q) & 
 \tilde L_{\phi}^{\Lam}(q) &  - \tilde L_x''^{\Lam}(q) \\
 0 & \tilde L_x'^{\Lam}(q) & 
 \tilde L_x''^{\Lam}(q) & \tilde L_c^{\Lam}(q)
 \end{array} \right) \; .
\end{equation}
The matrix elements $\tilde L^{\Lam}_{0j}$ and $\tilde L^{\Lam}_{j0}$ 
with $j=1,2,3$ vanish for symmetric form factors.
The other matrix elements are given by
\begin{eqnarray}
 \tilde L_c^{\Lam}(q) &=& \sum_p
 \left[ G^{\Lam}(p-\q2) G^{\Lam}(p+\q2) - 
 F^{\Lam}(p-\q2) F^{\Lam}(p+\q2) \right] f_c^2(p) \; ,
 \nonumber \\
 \tilde L_m^{\Lam}(q) &=& \sum_p
 \left[ G^{\Lam}(p-\q2) G^{\Lam}(p+\q2) + 
 F^{\Lam}(p-\q2) F^{\Lam}(p+\q2) \right] f_m^2(p) \; ,
 \nonumber \\
 \tilde L_a^{\Lam}(q) &=& - \sum_p
 \left\{ \Re\left[G^{\Lam}(p+\q2) G^{\Lam}(-p+\q2)\right] - 
 F^{\Lam}(p-\q2) F^{\Lam}(p+\q2)
 \right\} f_p^2(p) \; ,
\nonumber \\
 \tilde L_{\phi}^{\Lam}(q) &=& - \sum_p
 \left\{ \Re\left[G^{\Lam}(p+\q2) G^{\Lam}(-p+\q2)\right] + 
 F^{\Lam}(p-\q2) F^{\Lam}(p+\q2)
 \right\} f_p^2(p) \; ,
\nonumber \\
 \tilde L_p''^{\Lam}(q) &=& - \sum_p
 \Im\left[G^{\Lam}(p+\q2) G^{\Lam}(-p+\q2)\right] f_p^2(p) \; ,
\nonumber \\
 \tilde L_x'^{\Lam}(q) &=& 2 \sum_p
 \Re\left[G^{\Lam}(p-\q2) F^{\Lam}(p+\q2)\right] f_c(p) f_p(p) \; , 
\nonumber \\
 \tilde L_x''^{\Lam}(q) &=& 2 \sum_p
 \Im\left[G^{\Lam}(p-\q2) F^{\Lam}(p+\q2)\right] f_c(p) f_p(p) \; .
\end{eqnarray}

The system of linear equations Eq.~(\ref{rpa_Gamq}) can be 
solved explicitly. 
The magnetic coupling function is decoupled from the others, 
so that 
$M^{(0)}(q) + M^{\Lam}(q) = 
 \{ [M^{(0)}(q)]^{-1} - \tilde L_m^{\Lam}(q) \}^{-1}$.
Solving for the other coupling functions amounts to solving
a linear $3 \times 3$ system.
We do not write the explicit expressions here, but discuss
only the singularity structure of the coupling functions.
Singularities arise because the determinant
$d^{\Lam}(q) = \det[{\bf 1} - \tilde\bU(q)\tilde\bL^{\Lam}(q)]$ 
vanishes at $q=0$ for $\Delta_0 \to 0$, if $\Delta^{\Lam}(k)$ 
remains finite, that is, in case of spontaneous symmetry 
breaking.
For $q=0$, the explicit solution for the coupling functions 
and their behavior for $\Delta_0 \to 0$ was discussed in 
detail in Ref.~\onlinecite{eberlein10}.
For small $q \neq 0$, one can expand
$d(q) = d^{\Lam=0}(q) = d_0 + d_1 q_0^2 + d_2 \bq^2 + \dots$, 
where $d_0 \propto \Delta_0$ for small $\Delta_0$, while $d_1$ 
and $d_2$ remain finite for $\Delta_0 \to 0$.
Expanding all coefficients to leading order in $q_0$ and
$\bq$, one obtains the singular coupling functions
\begin{eqnarray} \label{rpa_sing}
 \Phi(q) &\propto& 
 - \frac{1}{d_0 + d_1 q_0^2 + d_2 \bq^2} \; ,
\nonumber \\
 P''(q) &\propto& 
 - \frac{q_0}{d_0 + d_1 q_0^2 + d_2 \bq^2} \; ,
\nonumber \\
 X''(q) &\propto& 
 - \frac{q_0}{d_0 + d_1 q_0^2 + d_2 \bq^2}
\end{eqnarray}
for $\Lam=0$.
The other coupling functions, $C^{\Lam}(q)$, $M^{\Lam}(q)$,
$A^{\Lam}(q)$, and the real part of $X^{\Lam}(q)$ remain finite 
for $\Lam \to 0$, $\Delta_0 \to 0$, $q \to 0$.

The divergence of the vertex in the phase fluctuation channel
represented by the coupling function $\Phi^{\Lam}(q)$ reflects
the Goldstone mode associated with the spontaneous breaking of
the $U(1)$ symmetry. The Goldstone theorem, which guarantees 
the existence of this mode, is obviously respected by the RPA.
A less familiar interesting result of the above calculation is the
divergence of the (3+1)-interaction represented by the coupling 
function $X^{\Lam}(q)$. At $q=0$ this interaction describes pair 
annihilation (or creation) combined with a forward scattering 
process.

%%%%%%%%%%%%%%%%%%%%%%%%%%%%%%%%%%%%%%%%%%%%%%%%%%%%%%%%%%%%%%%%%%

\section{Attractive Hubbard model}

In this section we compute the flow of the Nambu vertex and the 
gap function for the two-dimensional attractive Hubbard model as 
a prototype of a spin-singlet superfluid.

The Hubbard model describes interacting spin-$\frac{1}{2}$ 
lattice fermions with the Hamiltonian
\begin{equation}
 H = \sum_{\bi,\bj} t_{\bi\bj} c^{\dag}_{\bi\sg} c_{\bj\sg} +
 U \sum_{\bj} n_{\bj\up} n_{\bj\down} \; , 
\end{equation}
where $c^{\dag}_{\bi\sg}$ and $c_{\bi\sg}$ are creation and
annihilation operators for fermions with spin orientation $\sg$
on a lattice site $\bi$. For the attractive Hubbard model the
interaction parameter $U$ is negative.
The hopping matrix $t_{\bi\bj}$ is usually short-ranged.
We consider the case of nearest and next-to-nearest neighbor
hopping on a square lattice, with amplitudes $-t$ and $-t'$, 
respectively, yielding a dispersion relation of the form
\begin{equation}
 \eps(\bk) = - 2t \left( \cos k_x + \cos k_y \right)  
              - 4t' \cos k_x \cos k_y \; .
\end{equation}

The ground state of the attractive Hubbard model is a spin-singlet
$s$-wave superfluid at any filling factor.\cite{micnas90}
For $t'=0$ the superfluid order is degenerate with a charge density
wave at half-filling (only).
The attractive Hubbard model has been studied already in several
works both at zero and finite temperature by resummed perturbation
theory (mostly T-matrix),\cite{fresard92,pedersen97,letz98,keller99}
quantum Monte Carlo (QMC) methods,\cite{randeria92,santos94,singer96}
and dynamical mean-field theory.\cite{keller01,capone02}

%%%%%%%%%%%%%%%%%%%%%%%%%%%%%%%%%%%%%%%%%%%%%%%%%%%%%%%%%%%%%%%%%%%

\subsection{Regularization and counterterm}

The renormalization group flow is governed by the scale dependence
of the regularized bare propagator, which we choose to be of the
following form
\begin{equation} \label{G0reg}
 \left[ \bG_0^{\Lam}(k) \right]^{-1} = \left(  
 \begin{array}{cc}
 ik_0 - \xi(\bk) - \delta\xi^{\Lam}(\bk) + R^{\Lam}(k_0) &
 \Delta_0 \\
 \Delta_0 &
 ik_0 + \xi(\bk) + \delta\xi^{\Lam}(\bk) + R^{\Lam}(k_0)
 \end{array} \right) \; ,
\end{equation}
with $\xi(\bk) = \eps(\bk) - \mu$. The regulator function
\begin{equation} \label{regulator}
 R^{\Lam}(k_0) = i \, \sgn(k_0) \sqrt{k_0^2 + \Lam^2} - ik_0
\end{equation}
replaces frequencies $k_0$ with $|k_0| \ll \Lam$ by $\sgn(k_0) \Lam$ 
and thus regularizes the Fermi surface singularity of the bare 
fermionic propagator. 
The (real) bare gap $\Delta_0$ induces symmetry breaking and
regularizes the Goldstone mode singularity forming in the effective 
interaction below the critical scale $\Lam_c$. 
Instead of linking the flow of $\Delta_0$ to the fermionic cutoff 
scale $\Lam$ by defining a $\Lam$-dependent $\Delta_0^{\Lam}$, 
we found it more convenient to keep $\Delta_0$ fixed until 
$\Lam$ has decreased to $0$, and send $\Delta_0$ to zero 
afterwards. The equations for the latter flow are obtained 
simply by replacing $\Lam$-derivatives by derivatives with respect 
to $\Delta_0$. 
The counterterm $\delta\xi^{\Lam}(\bk)$ is linked to the normal
component of the self-energy by the condition
\begin{equation} \label{deltaxi}
 \frac{d}{d\Lam} \left[ \delta\xi^{\Lam}(\bk) + 
 \Sg^{\Lam}(0,\bk) \right] = 0 \; ,
\end{equation}
such that the Fermi surface remains fixed during the flow.
Since there is a contribution proportional to $\partial_{\Lam} 
\delta\xi^{\Lam}$ to the scale derivative of $\Sg^{\Lam}$, 
solving Eq.~(\ref{deltaxi}) for $\partial_{\Lam} \delta\xi^{\Lam}$
amounts to solving a linear integral equation.

%%%%%%%%%%%%%%%%%%%%%%%%%%%%%%%%%%%%%%%%%%%%%%%%%%%%%%%%%%%%%%%%%%%

\subsection{Parametrization}
\label{paramet}

We now specify the approximate parametrization of the self-energy
and the interaction vertex. 
Due to the local bare interaction and the pairing instability
occurring in the $s$-wave channel, the momentum dependence of
the normal and anomalous self-energy can be expected to be weak,
at least at weak coupling. Perturbation theory \cite{neumayr03} 
and previous functional RG calculations \cite{gersch08} showed
that this is indeed the case.
We therefore discard the momentum dependence of the self-energy,
keeping however the frequency dependence. The latter is
treated numerically by discretizing $\Sg^{\Lam}(k_0)$ and
$\Delta^{\Lam}(k_0)$ on a suitable grid.
The counterterm $\delta\xi^{\Lam}$ is then also momentum 
independent and can be interpreted as a shift of the chemical
potential.

The interaction vertex is fully described by the coupling
functions $C^{\Lam}_{kk'}(q)$ etc.\ introduced in Sec.~IV,
where singular momentum and frequency dependences have been
isolated in one variable $q$.
We now approximate these functions by the following ansatz:
\begin{eqnarray}
 C^{\Lam}_{kk'}(q) &=& 
 C^{\Lam}(q) \, g^{\Lam}_c(k_0) \, g^{\Lam}_c(k'_0) 
 \; , \nonumber \\
 M^{\Lam}_{kk'}(q) &=& 
 M^{\Lam}(q) \, g^{\Lam}_m(k_0) \, g^{\Lam}_m(k'_0)
 \; , \nonumber \\
 A^{\Lam}_{kk'}(q) &=& 
 A^{\Lam}(q) \, g^{\Lam}_a(k_0) \, g^{\Lam}_a(k'_0)
 \; , \nonumber \\
 \Phi^{\Lam}_{kk'}(q) &=& 
 \Phi^{\Lam}(q) \, g^{\Lam}_{\phi}(k_0) \, g^{\Lam}_{\phi}(k'_0)
 \; , \nonumber \\
 P''^{\Lam}_{kk'}(q) &=& 
 P''^{\Lam}(q) \, g^{\Lam}_{\phi}(k_0) \, g^{\Lam}_{\phi}(k'_0)
 \; , \nonumber \\
 X'^{\Lam}_{kk'}(q) &=& 
 X'^{\Lam}(q) \, g^{\Lam}_c(k_0) \, g^{\Lam}_a(k'_0)
 \; , \nonumber \\
 X''^{\Lam}_{kk'}(q) &=& 
 X''^{\Lam}(q) \, g^{\Lam}_c(k_0) \, g^{\Lam}_{\phi}(k'_0)
 \; .
\end{eqnarray}
The vertex thus assumes the form of a collection of boson-mediated 
interactions with bosonic propagators coupled to the fermions via 
fermion-boson vertices $g^{\Lam}$. 
The latter are normalized to one at zero frequency ($k_0 = 0$).
The momentum dependence on $\bk$ and $\bk'$ has thus been neglected, 
and the dependence on $k_0$ and $k'_0$ has been factorized. 
For the attractive Hubbard model, dependences on
$k$ and $k'$ are generated only at order $U^3$, and can thus
be expected to be weak at least at weak coupling. Neglecting 
the dependence on $\bk$ and $\bk'$ implies a restriction
to $s$-wave symmetry in charge, magnetic and pairing channels.
As a consequence, all triplet components vanish, such that
$A^{\Lam}_{kk'}(q) = A^{S,\Lam}_{kk'}(q)$,
$\Phi^{\Lam}_{kk'}(q) = \Phi^{S,\Lam}_{kk'}(q)$, and
$X^{\Lam}_{kk'}(q) = X^{S,\Lam}_{kk'}(q)$, and in the matrix
$\bV^{\rm PP,\Lam}$, Eq.~(\ref{VPP}), only four elements are
non-zero.
Compared to an exact decomposition of the coupling functions as
in Eqs.~(\ref{decomp}) and (\ref{decompX}), the sum over basis
functions is replaced by just one term in the above ansatz.
Due to time-reversal and exchange symmetries there is no 
contribution to $W''^{\Lam}_{kk'}(q)$ of that form.
We have allowed for four distinct fermion-boson vertices
$g^{\Lam}_c$, $g^{\Lam}_m$, $g^{\Lam}_a$, and $g^{\Lam}_{\phi}$.
The factorization of the coupling functions is similar to the
factorization Eq.~(\ref{rpa_fact}) obtained for separable 
interactions in RPA. Instead of parametrizing the fermion-boson
vertices in the pairing channel by a single function $g^{\Lam}_p$,
we now distinguish between $g^{\Lam}_a$ and $g^{\Lam}_{\phi}$. 
It turns out that they differ only slightly.
The imaginary part of the pairing coupling function 
$P''^{\Lam}_{kk'}(q)$ has little impact on the flow. Instead of
introducing another fermion-boson vertex for that quantity, we 
approximate its dependence on $k_0$ and $k'_0$ by $g^{\Lam}_{\phi}$.
The frequency dependence of the fermion-boson vertices 
$g^{\Lam}(k_0)$ is treated numerically by discretization.

The parametrization of the ``boson propagators'' $C^{\Lam}(q)$ etc.\ 
requires special care, to capture the singularities.
We first consider the amplitude and phase channel.
For small $q$ the functions $A^{\Lam}(q)$ and $\Phi^{\Lam}(q)$ 
behave as 
$[A^{\Lam}(q)]^{-1} = 
 - m^{\Lam}_a - Z^{\Lam}_a \bq^2 - \bar Z^{\Lam}_a q_0^2 + \dots$ 
and
$[\Phi^{\Lam}(q)]^{-1} = - m^{\Lam}_{\phi} 
 - Z^{\Lam}_{\phi} \bq^2 - \bar Z^{\Lam}_{\phi} q_0^2 + \dots$, 
where
$m^{\Lam}_{\phi} \to 0$ for $\Lam < \Lam_c$, $\Delta_0 \to 0$.
Actually the regulator function can also generate contributions 
of order $|q_0|$, which disappear again as $\Lam \to 0$.
To deal with this technical complication, and to achieve an
accurate parametrization also at larger values of $q_0$ and $\bq$,
we parametrize $A^{\Lam}$ and $\Phi^{\Lam}$ by two scale-dependent
functions,
\begin{eqnarray} \label{Aq}
 \left[ A^{\Lam}(q) \right]^{-1} &=& 
 - m^{\Lam}_a(q_0) - e^{\Lam}_a(\bq) \; ,
 \nonumber \\
 \left[ \Phi^{\Lam}(q) \right]^{-1} &=&
 - m^{\Lam}_{\phi}(q_0) - e^{\Lam}_{\phi}(\bq) \; ,
\end{eqnarray}
where $e^{\Lam}_a(\b0) = e^{\Lam}_{\phi}(\b0) = 0$.
The (discretized) momentum and frequency dependences of these 
functions are determined from the flow.
The above ansatz with functions of one ($q_0$) and two ($\bq = (q_x,q_y)$)
variables reduces the numerical effort compared to a discretization
of an arbitrary function of $q =(q_0,q_x,q_y)$.
Tests within RPA indicate that it describes the full functions
sufficiently well. In particular, the behavior at small $q_0$ and 
small $\bq$ is captured correctly.
The imaginary parts $P''^{\Lam}(q)$ and $X''^{\Lam}(q)$ are odd 
functions of $q_0$. This and the expected singularity structure 
[see Eq.~(\ref{rpa_sing})] motivate the ansatz
\begin{eqnarray} \label{Pq}
 P''^{\Lam}(q) &=& 
 - \frac{q_0}{m^{\Lam}_{p''}(q_0) + e^{\Lam}_{p''}(\bq)} \; ,
 \nonumber \\
 X''^{\Lam}(q) &=& 
 - \frac{q_0}{m^{\Lam}_{x''}(q_0) + e^{\Lam}_{x''}(\bq)} \; .
\end{eqnarray}

The parametrization of $C^{\Lam}(q)$, $M^{\Lam}(q)$ and $X'^{\Lam}(q)$
is slightly more complicated, because at small $\bq$ these functions 
cannot be represented as a sum of a frequency and a momentum dependent 
function. We therefore distinguish the cases $|\bq| < q_{\rm max}$ and
$|\bq| > q_{\rm max}$ with a suitably chosen $q_{\rm max}$.
For $|\bq| > q_{\rm max}$ we make an additive ansatz analogous to 
Eq.~(\ref{Aq}),
\begin{eqnarray} \label{Cq1}
 \left[ C^{\Lam}(q) \right]^{-1} &=& 
 - m^{\Lam}_c(q_0) - e^{\Lam}_c(\bq) \; ,
 \nonumber \\
 \left[ M^{\Lam}(q) \right]^{-1} &=&
 - m^{\Lam}_m(q_0) - e^{\Lam}_m(\bq) \; ,
 \nonumber \\
 \left[ X'^{\Lam}(q) \right]^{-1} &=& 
 - m^{\Lam}_{x'}(q_0) - e^{\Lam}_{x'}(\bq) \; .
\end{eqnarray}
For small $\bq$, the $\bq$-dependence is increasingly isotropic,
except for the special case where the Fermi surface touches van
Hove points (which we exclude).
Hence, for $|\bq| < q_{\rm max}$ we approximate the momentum 
dependence as isotropic,
\begin{eqnarray} \label{Cq2}
 C^{\Lam}(q) &=& C^{\Lam}(q_0,|\bq|) \; , \nonumber \\
 M^{\Lam}(q) &=& M^{\Lam}(q_0,|\bq|) \; , \nonumber \\
 X'^{\Lam}(q) &=& X'^{\Lam}(q_0,|\bq|) \; ,
\end{eqnarray}
reducing the number of variables again to two.
To avoid a discontinuity at $q_{\rm max}$, we connect the two regimes
in momentum space by a smooth partition of unity instead of step 
functions.

%%%%%%%%%%%%%%%%%%%%%%%%%%%%%%%%%%%%%%%%%%%%%%%%%%%%%%%%%%%%%%%%%%%%

\subsection{Flow equations}

The flow equations for the scale-dependent functions parametrizing
the self-energy and interaction vertex are obtained by projecting
the flow equations for the self-energy and the coupling functions
on the simplified ansatz.
Dependences on the fermionic momenta $\bk$, $\bk'$ generated by 
the flow are eliminated by a Fermi surface average (but not the
$\bq$-dependence, of course). 
This corresponds to keeping only the leading (in power-counting) 
term in an expansion around the Fermi surface,
and averaging over the momentum dependence along the Fermi surface,
which is in line with the pure $s$-wave ansatz for the interactions.

For the self-energy, we project on the momentum-independent ansatz
by averaging the flow Eq.~(\ref{floweq_Sg}) over the Fermi surface
as follows:
\begin{equation}
 \frac{d}{d\Lam} \bSg^{\Lam}(k_0) =
 \bra {\bf rhs}^{\Lam}(k_0,\bk) \ket_{\bk \in {\rm FS}} \; ,
\end{equation}
where ${\bf rhs}^{\Lam}(k_0,\bk)$ stands for the right hand side 
of the flow equation (in Nambu matrix form), and 
$\bra \dots \ket_{\bk \in {\rm FS}}$ denotes a Fermi surface
average.
Momentum dependences of the self-energy perpendicular to the Fermi 
surface are marginal in power-counting \cite{shankar94} 
and lead to a (finite) renormalization of the Fermi velocity. 
However, they are quantitatively small in the attractive Hubbard 
model, at least at weak coupling and away from van Hove points, 
and have thus little influence on our results. 

The flow equations for the coupling functions $C^{\Lam}_{kk'}(q),\dots,
X^{\Lam}_{kk'}(q)$ were derived in Sec.~IV. Inserting the ansatz for 
the interaction vertex on the right hand side of these equations
yields several terms which can be represented by Feynman diagrams
of the form plotted in Fig.~5.
We recall that the point-like vertex represents the bare (here 
Hubbard) interaction, the wiggly line any coupling function 
contributing to $\bV^{\rm PH,\Lam}$, and the double line coupling
functions appearing in $\bV^{\rm PP,\Lam}$ (only $M^{\Lam}_{kk'}(q)$
in the absence of triplet terms). Note that the terms in Fig.~6
are redundant since the complete set of coupling functions is
already captured by $\bV^{\rm PH,\Lam}$.
The Nambu index sums on the right hand side of the flow equation
for the coupling functions can be transformed to a more convenient
form by representing the vertex and the propagator product in
the Pauli matrix basis defined in Appendix A and used already
in Sec.~V.

Some of the contributions, having the form of vertex corrections
or box diagrams, generate dependences on $\bk$ and $\bk'$ which
are not allowed for in our ansatz.
These dependences are projected out by a symmetrized Fermi surface
average. We discuss the procedure for the charge coupling function
as a prototypical example, which can be extended directly to all
other cases. The flow of the projected charge coupling function 
is given by
\begin{eqnarray} \label{floweqCproj}
 \frac{d}{d\Lam} \left[ C^{\Lam}(q) \, 
 g^{\Lam}_c(k_0) \, g^{\Lam}_c(k'_0) \right] &=&
 \bra {\rm rhs}^{\Lam}(k,k';q) 
 \ket_{\bk\pm\frac{\bq}{2},\bk'\pm\frac{\bq}{2} \in {\rm FS}} 
 \nonumber \\ & \equiv &
 \overline{\rm rhs}^{\Lam}(k_0,k'_0;q) \; ,
\end{eqnarray}
where ${\rm rhs}^{\Lam}(k,k';q)$ denotes the complete right hand
side of the flow equation for $C^{\Lam}_{kk'}$, Eq.~(\ref{floweqC}),
and
\begin{equation}
 \bra \dots \ket_{\bk\pm\frac{\bq}{2},\bk'\pm\frac{\bq}{2} 
 \in {\rm FS}} =
 \frac{1}{4} \sum_{\eps,\eps' = \pm 1}
 \bra \dots \ket_{\bk+\eps\frac{\bq}{2},\bk'+\eps'\frac{\bq}{2} 
 \in {\rm FS}}
\end{equation}
is a symmetrized Fermi surface average. The latter averages
the four possible ways of integrating $\bk$ and $\bk'$ under
the constraint that two of the four external momenta
$\bk\pm\frac{\bq}{2}$ and $\bk'\pm\frac{\bq}{2}$ of the vertex 
lie on the Fermi surface.
For $\bq = \b0$, corresponding to the forward scattering and
Cooper channels, this becomes a Fermi surface average with 
all four momenta on the Fermi surface.
For $\bq \neq \b0$, the set of momenta $\bk$ satisfying both 
$\bk + \frac{\bq}{2} \in {\rm FS}$ and $\bk - \frac{\bq}{2} 
\in {\rm FS}$ is limited to few points, except for special 
nesting vectors in case of a nested Fermi surface.
The Fermi surface average picks up the $s$-wave component of
the dominant processes near the Fermi surface.
Indeed, the momentum dependence perpendicular to the Fermi 
surface is irrelevant in power-counting.\cite{shankar94}
Note, however, that we do not discard the dependence on $\bq$.
That dependence becomes important due to the formation of
singularities, which invalidate the weak-coupling 
power-counting.

The projection on the form factors in the channel decomposition 
could also be carried out by integration over the entire
Brillouin zone.\cite{husemann09,husemann12}
However, for our simple ansatz with only one (momentum-independent) 
form factor it is better to approximate the vertex by its Fermi
surface average instead of a Brillouin zone average, to capture 
the dominant contributions at low energy scales. We checked this
in some test cases by explicit comparison of different projection
procedures.

The flow equation for the bosonic propagator can be extracted 
from Eq.~(\ref{floweqCproj}) by setting $k_0 = k'_0 = 0$.
Since $g^{\Lam}_c(0) = 1$ is independent of $\Lam$, one obtains
\begin{equation}
 \frac{d}{d\Lam} C^{\Lam}(q) = 
 \overline{\rm rhs}^{\Lam}(0,0;q) \; .
\end{equation}
The functions $m_c^{\Lam}(q_0)$ and $e_c^{\Lam}(\bq)$ parametrizing 
$C^{\Lam}(q)$ for $|\bq| > q_{\rm max}$ are extracted
by evaluating $[C^{\Lam}(q)]^{-1}$ at a fixed momentum $\bq^*$ 
as a function of $q_0$, and at fixed frequency $q_0 = 0$ as a
function of $\bq$, respectively.
For $\bq^*$ we choose a momentum where $C^{\Lam}(q)$ is peaked,
where it yields the largest contribution. In the charge and
magnetic channel this happens typically at finite momenta 
connecting antipodal Fermi points ($2k_F$-type momenta).

The product rule for differentiation applied to the left
hand side of Eq.~(\ref{floweqCproj}) at $k'_0 = 0$ yields
the flow equation for the fermion-boson vertex,
\begin{equation} \label{vertexproj}
 \frac{d}{d\Lam} g^{\Lam}_c(k_0) = \frac{1}{C^{\Lam}(q)}
 \left[ \overline{\rm rhs}^{\Lam}(k_0,0;q) - 
 g^{\Lam}_c(k_0) \, 
 \overline{\rm rhs}^{\Lam}(0,0;q) \right]_{q = (0,\bq^*)} \; .
\end{equation}
The flow equations for the other coupling functions 
$M^{\Lam}_{kk'}(q)$ etc.\ are projected on the ansatz in the 
same way. The flow of the fermion-boson vertices in the pairing
channel $g^{\Lam}_a(k_0)$ and $g^{\Lam}_{\phi}(k_0)$ is determined
as in Eq.~(\ref{vertexproj}), with $\bq^* = \b0$.

The initial conditions for the flow at $\Lam_0 = \infty$ are as 
follows. For the self-energy, counterterm and gap function the
flow starts at $\Sg^{\Lam_0} = 0$, $\delta\xi^{\Lam_0} = 0$, and
$\Delta^{\Lam_0} = \Delta_0$. The coupling functions are initially
zero, and the fermion-boson vertices are equal to one.
Note that the coupling functions do not include the bare 
interaction.
In a numerical evaluation, the flow starts at a large finite 
$\Lam_0$. The self-energy receives a tadpole contribution of
order one in the flow from $\Lam_0 = \infty$ to an arbitrarily
large finite $\Lam_0$, yielding $\Sg^{\Lam_0} = U/2$ with 
corrections of order $\Lam_0^{-1}$, and correspondingly 
$\delta\xi^{\Lam_0} = - U/2$.
The error of order $\Lam_0^{-1}$ made by starting the flow at
a (large) finite cutoff can be significantly reduced by using
perturbative results at $\Lam_0$ as initial conditions instead
of the initial values at $\Lam_0 = \infty$.
The coupling functions are then non-zero from the beginning
such that Eq.~(\ref{vertexproj}) is well defined at $\Lam_0$.

We conclude this section with a few words on numerical aspects.
More details can be found in Ref.~\onlinecite{eberlein13}.
Momentum and frequency dependences were discretized on 
non-equidistant grids such that the resolution is higher at
smaller frequencies and momenta.
The positive frequency axis and radial momentum dependences 
were discretized by around 30 points, and angular momentum
dependences by 6 angles per quadrant in the Brillouin zone.
The functional flow equations were thus replaced by a system
of around 2000 non-linear ordinary differential equations
with three-dimensional loop integrals on the right hand 
sides. The integrals were performed with an adaptive
integration algorithm and the integration of the flow was
performed with a third-order Runge-Kutta routine.
Depending on parameters, the computation of a flow required
between a day and a week on 20 CPU cores.

%%%%%%%%%%%%%%%%%%%%%%%%%%%%%%%%%%%%%%%%%%%%%%%%%%%%%%%%%%%%%%%%%%

\subsection{Results}

We now present results for the effective interactions, the normal
self-energy, and the gap function as obtained from a numerical
solution of the flow equations. 
Most of the numerical results are obtained for a small fixed 
external pairing field $\Delta_0$ chosen two to three orders of
magnitude below the mean-field gap $\Delta_{\rm MF}$, but we
also discuss some flows where $\Delta_0$ scales toward zero
after the fermionic cutoff has reached $\Lam = 0$.
The Ward identity following from global charge conservation is
enforced at zero frequency by projecting the flow, if not stated 
otherwise (for details, see Sec.~\ref{wardidentity}).
Bare interaction strengths are chosen in the weak to moderate
coupling range $|U|/t = 1 - 4$. 
%Next-nearest neighbor hoppings are $t'/t = -0.1$ or $0$, and the 
%particle density is $n=0.5$ (quarter-filling) or $n=0.78$.
In the following we set the nearest-neighbor hopping amplitude 
$t=1$, that is, all quantities with dimension of energy are in 
units of $t$.

%%%%%%%%%%%%%%%%%%%%%%%%%%%%%%%%%%%%%%%%%%%%%%%%%%%%%%%%%%%%%%%%%%

\subsubsection{Effective interactions}

We begin with results for the coupling functions, which describe
the various effective interaction channels contributing to the
the Nambu vertex. With our sign conventions negative coupling 
functions correspond to attractive effective interactions in 
the respective channel.

The flow of effective interactions in the pairing channel is 
qualitatively similar to the one in RPA (see Sec.~V). However,
the critical scale and the size of the coupling functions is 
reduced by fluctuations.
Typical flows for the amplitude and phase couplings at $q=0$ are
shown in Fig.~7 for various choices of the external pairing field
$\Delta_0$.
\begin{figure}[tb]
\centerline{\includegraphics[width=8cm]{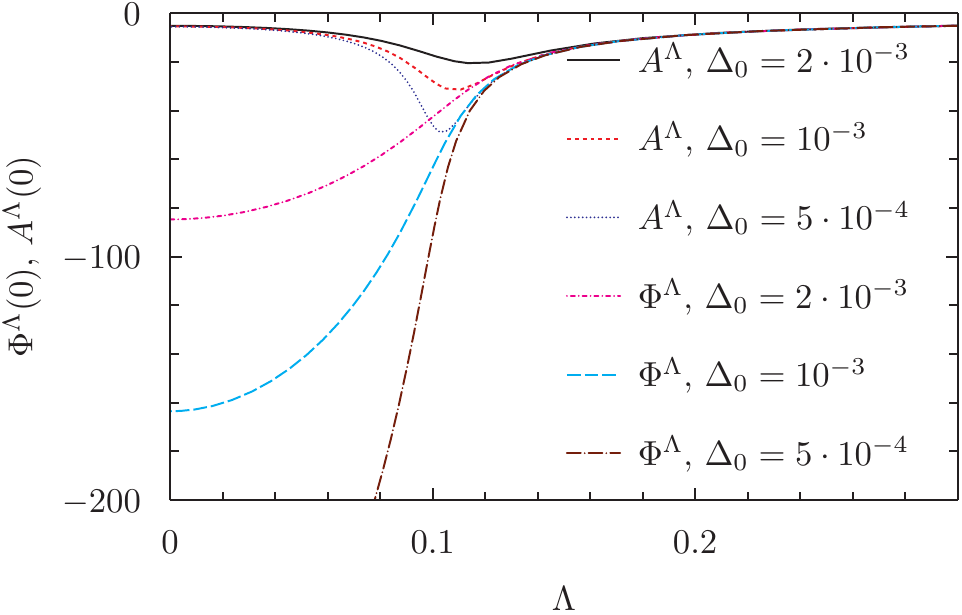}}
\caption{(Color online)
Scale dependence of the amplitude and phase couplings
$A^{\Lam}$ and $\Phi^{\Lam}$ at $q=0$ for various choices 
of the external pairing field $\Delta_0$. The Hubbard model 
parameters are $t'=-0.1$, $U=-2$, $n=0.5$ (quarter-filling).}
\end{figure}
For $U=-2$ stable flows without artificial singularities could be 
performed for external pairing fields as small as three orders of 
magnitude below the size of the mean-field gap $\Delta_{\rm MF}$,
with a final phase coupling $\Phi^{\Lam=0}(0)$ proportional to 
$\Delta_0^{-1}$ within numerical accuracy, as dictated by the
Ward identity.
The amplitude coupling $A^{\Lam}(0)$ has a peak around $\Lam_c$,
whose size increases strongly upon reducing $\Delta_0$, while it
reaches a finite value with a negligible dependence on the external 
pairing field at the end of the flow.

The momentum and frequency dependence of $A^{\Lam}(q)$ and
$\Phi^{\Lam}(q)$ around $q=0$ is shown for various choices of
$\Lam$ in Fig.~8.
\begin{figure}[tb]
\centerline{\includegraphics[width=6cm]{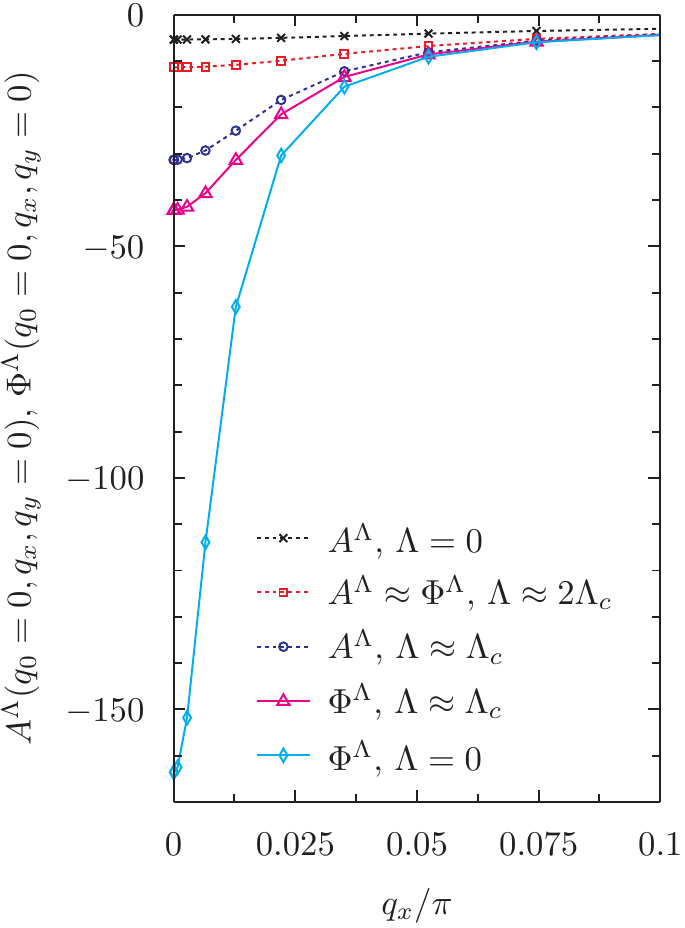} \hskip 5mm
 \includegraphics[width=6cm]{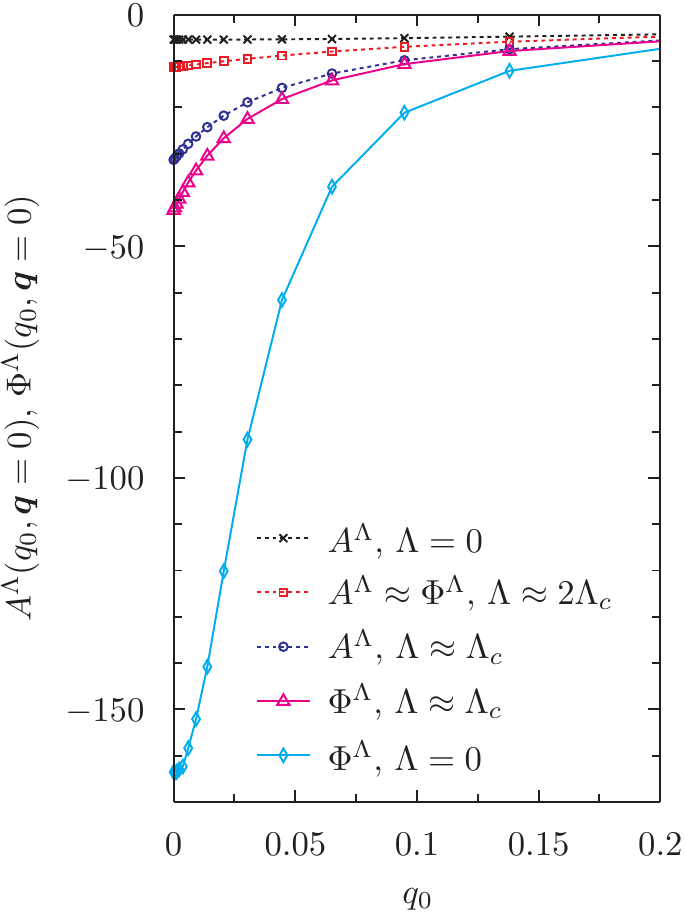}}
\caption{(Color online)
Momentum dependence along the $q_x$-axis (left) and frequency 
dependence (right) of the amplitude and phase couplings 
$A^{\Lam}(q)$ and $\Phi^{\Lam}(q)$ for small momenta and 
frequencies at various stages of the flow. 
The Hubbard model parameters are the same as in Fig.~7 and
$\Delta_0 = 10^{-3}$.}
\end{figure}
For small momenta, the coupling functions are isotropic functions 
of $\bq$ with a momentum dependence proportional to $\bq^2$,
for both finite $\Lam$ and $\Lam = 0$.
The frequency dependence is linear for small $q_0$ at $\Lam > 0$,
but essentially quadratic for $\Lam = 0$. The linear behavior at 
$\Lam > 0$ is caused by the frequency dependent regulator, 
Eq.~(\ref{regulator}), and thus disappears once the regulator has 
scaled to zero.
The amplitude coupling $A^{\Lam=0}(q_0,\b0)$ exhibits a tiny
dip at $q_0 = 0$.
Overall, the qualitative momentum and frequency dependences of 
the coupling functions in the pairing channel do not deviate 
significantly from the behavior in RPA.
This is also true for the imaginary part of the pairing coupling
$P''^{\Lam}(q)$ and the imaginary part of the anomalous
(3+1)-coupling $X''^{\Lam}(q)$, whose singular behavior at
small momenta and frequencies is well described by
\begin{equation}
 X''^{\Lam=0}(q) \propto P''^{\Lam=0}(q) \propto
 - \frac{q_0}{\Delta_0 + aq_0^2 + b\bq^2} \; ,
\end{equation}
where $a,b$ are positive constants.

The charge coupling function $C^{\Lam}(q)$ is generally
negative at all stages of the flow. It thus renormalizes 
the bare attraction in the charge channel given by $U$ to 
an enhanced total attractive interaction $U + 2C^{\Lam}(q)$.
This effect is captured already in RPA.
The enhancement is usually small. However, for densities
near half-filling and small values of $t'$ it becomes large 
at $q = (0,\bQ)$ with $\bQ = (\pi,\pi)$.
For $t'=0$ and half-filling, $U + 2C^{\Lam}(0,\bQ)$ is 
degenerate with the pairing interaction $U + P^{\Lam}(0,\b0)$,
reflecting the degeneracy of superfluidity with charge
density wave order due to a particle-hole symmetry in this 
special case.\cite{micnas90}
In Fig.~9 we show the momentum dependence of the charge
coupling function in the static limit $q_0 = 0$ at the end of
the flow ($\Lam = 0$) for two distinct choices of $t'$ and $n$. 
\begin{figure}[tb]
\centerline{\includegraphics[width=7cm]{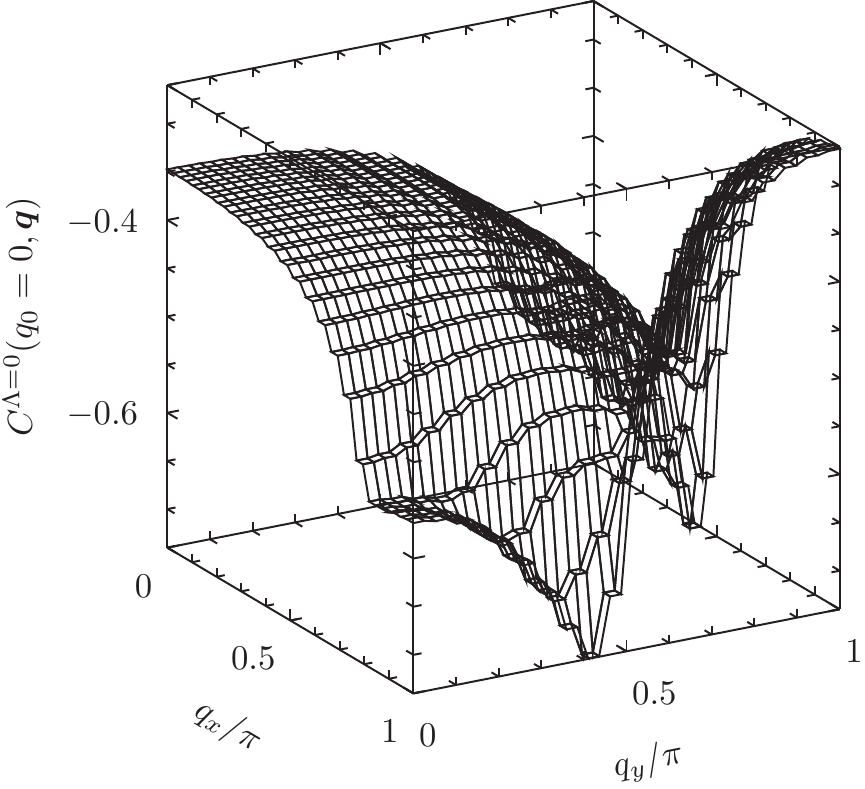} \hskip 5mm
 \includegraphics[width=7cm]{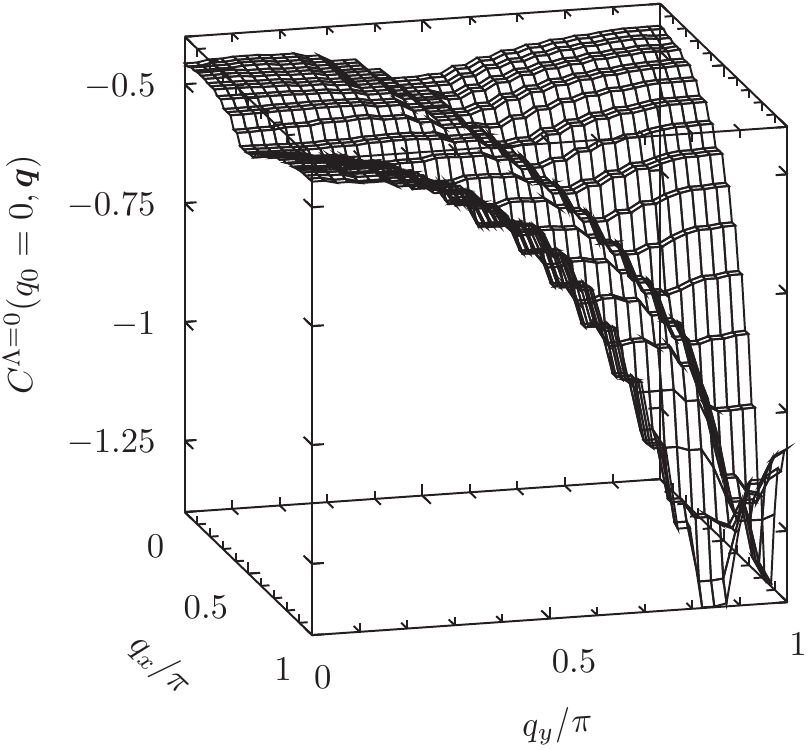}}
\caption{Momentum dependence of the static charge coupling 
function $C^{\Lam=0}(0,\bq)$ at the end of the flow, for 
$t'=-0.1$, $n=0.5$ (left) and $t'=0$, $n=0.78$ (right).
The Hubbard interaction is $U = -2$ in both cases.}
\end{figure}
The function exhibits pronounced peaks at incommensurate 
momenta situated at the Brillouin zone boundary. These peaks
are present already in the bare polarization function 
(particle-hole bubble).\cite{holder12}
They move toward $(\pi,\pi)$ and increase upon approaching 
half-filling for $t'=0$.
As a function of frequency, the size of $C^{\Lam}(q)$ decays 
monotonically upon increasing $|q_0|$.

The magnetic coupling function $M^{\Lam}(q)$ receives contributions
beyond RPA which change its behavior qualitatively.
In Fig.~10 we show its momentum dependence in the static limit
$q_0 = 0$ at the end of the flow for the same choices of $t'$ and
$n$ as in Fig.~9. 
\begin{figure}[tb]
\centerline{\includegraphics[width=7cm]{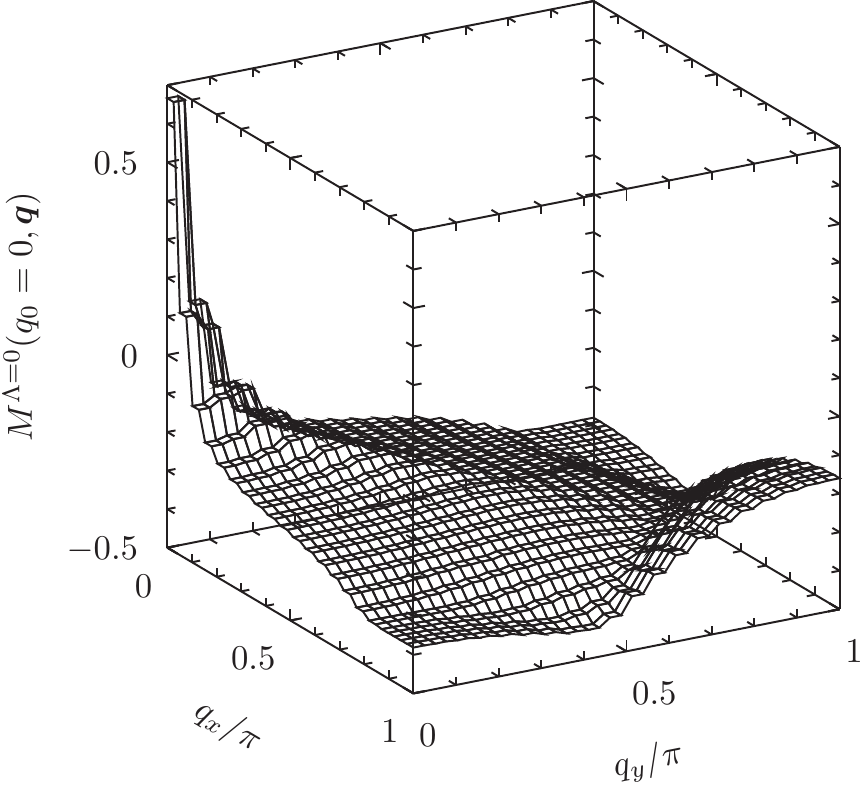} \hskip 5mm
 \includegraphics[width=7cm]{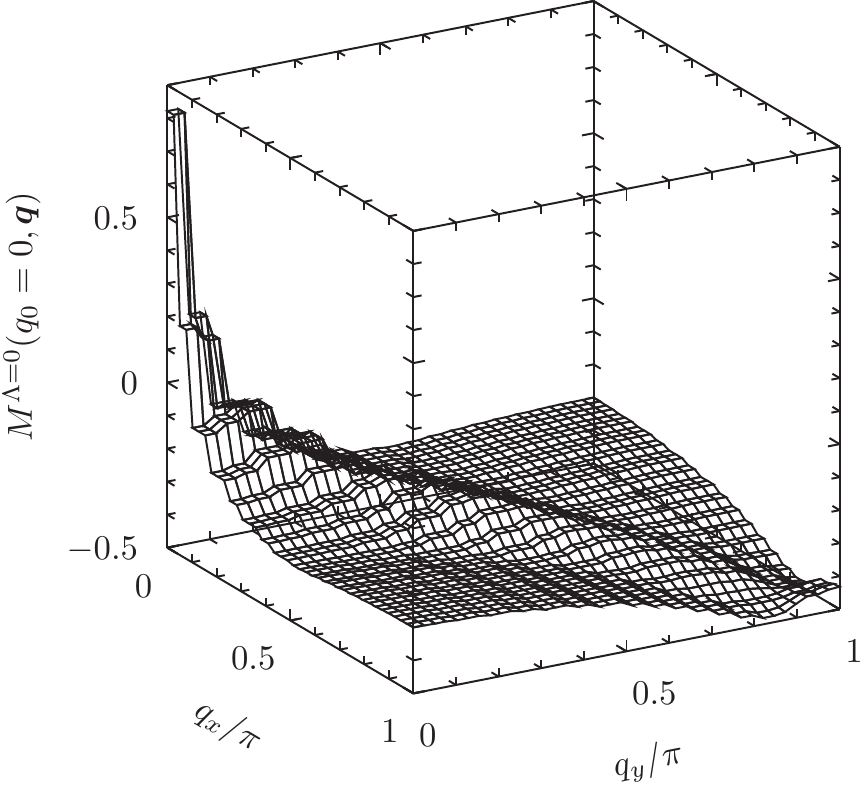}}
\caption{Momentum dependence of the static magnetic coupling 
function $M^{\Lam=0}(0,\bq)$ at the end of the flow, for 
$t'=-0.1$, $n=0.5$ (left) and $t'=0$, $n=0.78$ (right).
The Hubbard interaction is $U = -2$ in both cases.}
\end{figure}
The coupling function is negative in most of the Brillouin zone, 
but it develops a pronounced positive peak for small momenta $\bq$.
This peak is a pure fluctuation effect. In RPA, $M^{\Lam=0}(0,\b0)$
vanishes due to the pairing gap.
Since the amplitude of the coupling function is small, the total 
interaction in the magnetic channel $-U + 2M^{\Lam=0}(q)$ is 
dominated by the bare Hubbard interaction and remains positive 
for all momenta.
In Fig.~11 the frequency dependence of $M^{\Lam}(q_0,\b0)$ is
shown at various stages of the flow. 
\begin{figure}[tb]
\centerline{\includegraphics[width=6cm]{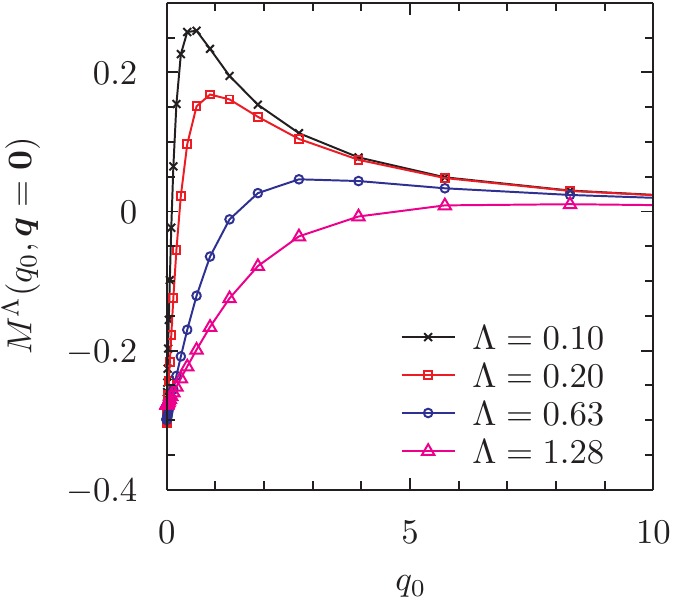} \hskip 5mm
 \includegraphics[width=6.2cm]{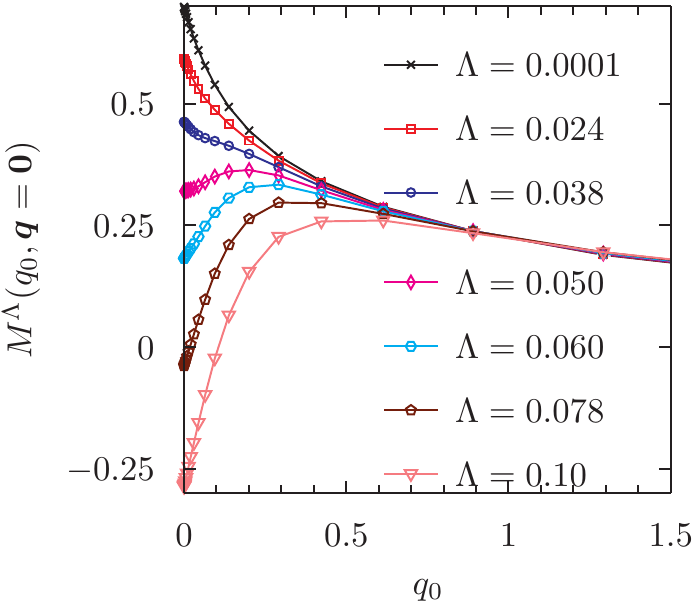}}
\caption{(Color online)
 Frequency dependence of the magnetic coupling function
 $M^{\Lam}(q_0,\b0)$ at various stages of the flow above (left) 
 and below (right) the critical scale for pairing. 
 The model parameters are $t'=-0.1$, $U=-2$, and $n=0.5$.}
\end{figure}
The positive peak at $q_0 = 0$ develops at and below the critical
scale $\Lam_c$ and is foreshadowed by a finite frequency peak for
$\Lam$ near $\Lam_c$.
The flow is non-monotonic and $M^{\Lam}(q_0,\b0)$ exhibits a sign 
change for small $q_0$, but eventually $M^{\Lam=0}(q_0,\b0)$ is 
positive for all frequencies.
A similar sign change at finite $q_0$ and a pronounced finite
frequency peak has been observed previously in the charge 
coupling function for the repulsive Hubbard model in the 
symmetric regime ($\Lam > \Lam_c$).\cite{husemann12,giering12}

The real part of the anomalous (3+1)-coupling function 
$X'^{\Lam}(q)$ is relatively small. Its singularity at the
critical scale $\Lam_c$ is considerably broadened by 
fluctuations (beyond RPA). Nevertheless, its influence on
the flow of the self-energy and the other coupling functions 
is important. Neglecting the (3+1)-coupling would lead to 
artifacts like non-monotonic flows of $\Phi^{\Lam}(0)$ even 
for small interactions $U$.
While the imaginary part of $X^{\Lam}(q)$ depends strongly
on $\Delta_0$ for small $q$ and $\Lam$, the real part does 
not.
In Fig.~12 we plot the momentum dependence of the static
(3+1)-coupling function at the end of the flow for the same 
choices of $t'$ and $n$ as in Figs.~9 and 10. Note that the 
imaginary part of the static (3+1)-coupling function 
vanishes, that is, $X^{\Lam}(0,\bq)$ is real.
\begin{figure}[tb]
\centerline{\includegraphics[width=7cm]{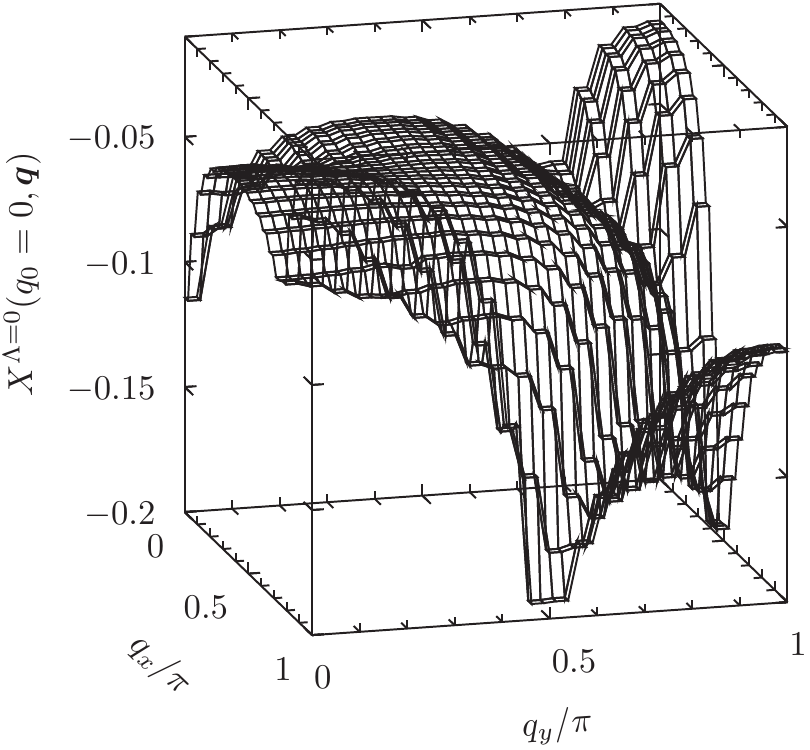} \hskip 5mm
 \includegraphics[width=7cm]{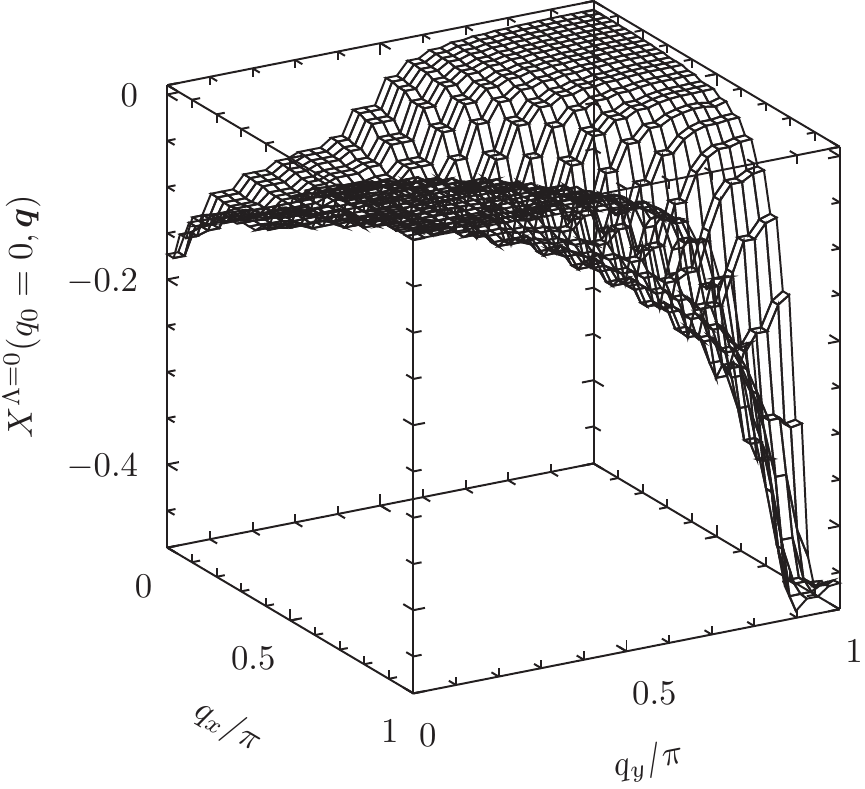}}
\caption{Momentum dependence of the static (3+1)-coupling 
function $X^{\Lam=0}(0,\bq)$ at the end of the flow, for 
$t'=-0.1$, $n=0.5$ (left) and $t'=0$, $n=0.78$ (right).
The Hubbard interaction is $U = -2$ in both cases.}
\end{figure}

We now turn to the fermion-boson vertices, whose frequency
dependence is plotted in Fig.~13.  
\begin{figure}[tb]
\centerline{\includegraphics[width=7.5cm]{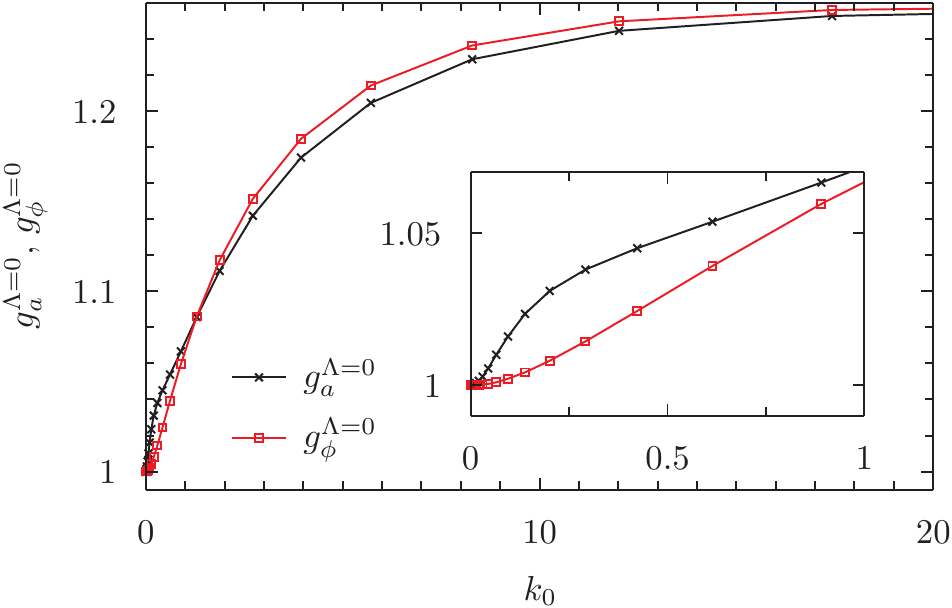} \hskip 5mm
 \includegraphics[width=7.5cm]{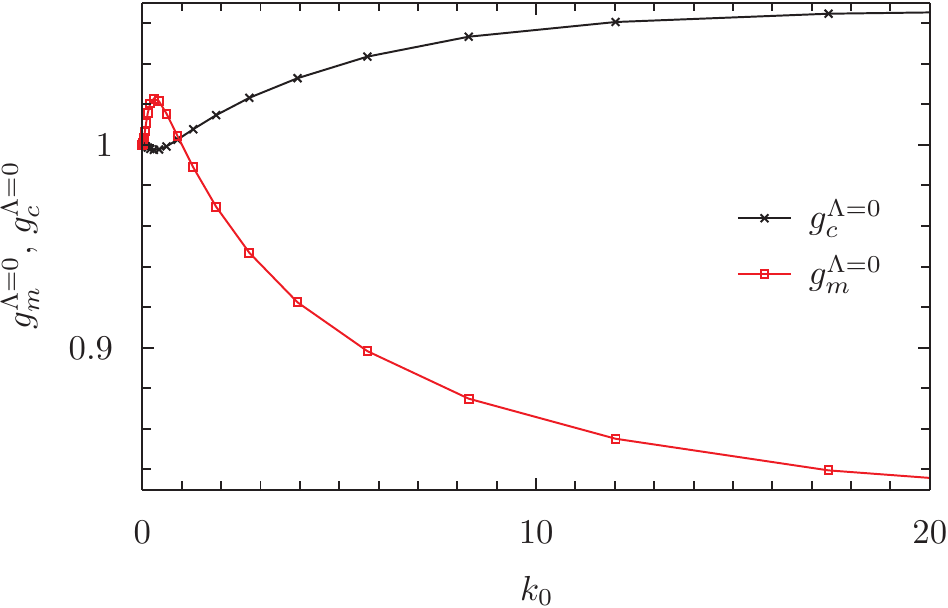}}
\caption{(Color online)
 Frequency dependence of the fermion-boson vertices
 at the end of the flow ($\Lam=0$).
 Left: amplitude and phase vertices. 
 Right: charge and magnetic vertices.
 The model parameters are $t'=-0.1$, $U=-2$, and $n=0.5$.}
\end{figure}
Note that the vertices are even functions of $k_0$ which are
normalized to one at $k_0 = 0$ by definition.
The frequency dependence of the vertices is quite weak.
However, the frequency dependence of the vertices in the 
pairing channel contributes significantly to the frequency 
dependence of the gap function $\Delta^{\Lam}(k_0)$ and 
also to the flow of $\Phi^{\Lam}$. The normal self-energy
and the other coupling functions are only weakly affected
by the frequency dependence of the fermion-boson vertices.
The magnetic vertex exhibits a small peak at low frequencies
which develops at scales $\Lam < \Lam_c$ and is therefore 
related to pairing fluctuations.

%%%%%%%%%%%%%%%%%%%%%%%%%%%%%%%%%%%%%%%%%%%%%%%%%%%%%%%%%%%%%%%%%%%%%%

\subsubsection{Normal self-energy and gap function}
\label{selfgap}

At weak to moderate interactions the ground state of the attractive
Hubbard model is superfluid with Cooper pairs made of weakly 
renormalized quasi particles. 
Quasi particle renormalization occurs already at scales above the 
pairing scale $\Lam_c$ and is described by the normal self-energy.
The momentum dependence of the self-energy is weak for the choice of
parameters considered in this work and we will present only results
for the Fermi surface average 
$\Sg^{\Lam}(k_0) = \bra \Sg^{\Lam}(k_0,\bk) \ket_{\bk \in {\rm FS}} \,$.
In Fig.~14 we show results for the imaginary part of $\Sg^{\Lam}(k_0)$
as a function of frequency at the end of the flow ($\Lam=0$).
\begin{figure}[tb]
\centerline{\includegraphics[width=8cm]{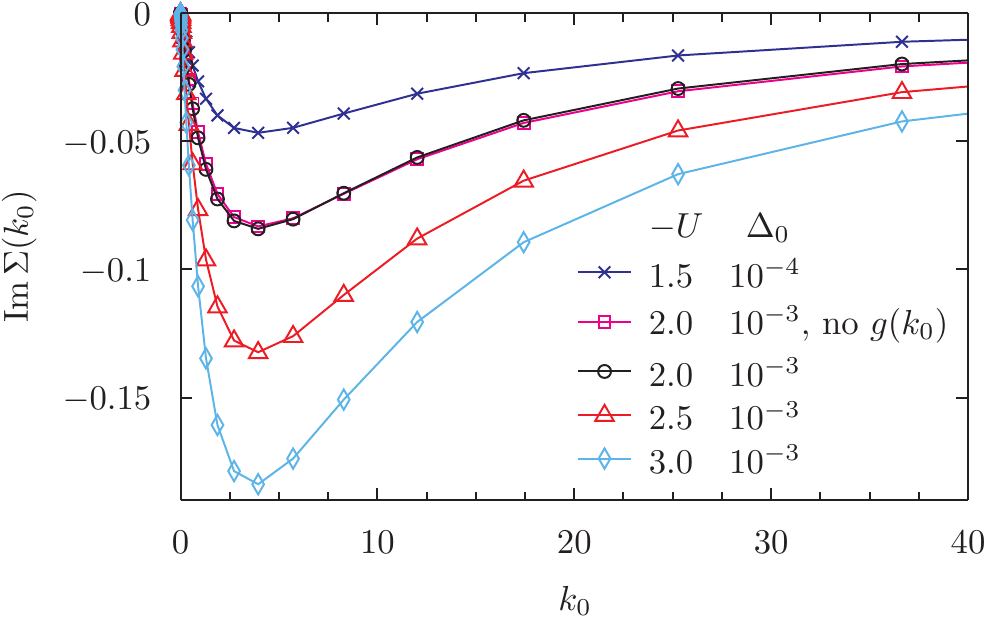}}
\caption{(Color online)
 Frequency dependence of the imaginary part of the
 normal self-energy for $t'=-0.1$, $n=0.5$ (quarter-filling)
 and various choices of $U$ at the end of the flow. 
 The result labeled as ``no $g(k_0)$'' is obtained with constant 
 (frequency-independent) fermion-boson vertices.}
\end{figure}
We plot only the positive frequency axis since $\Im\Sg^{\Lam}(-k_0) = 
- \Im\Sg^{\Lam}(k_0)$.
The real part (not plotted) of $\Sg^{\Lam}(k_0)$ is an even function of 
$k_0$ with a negative peak at $k_0=0$ that decays monotonically to the
Hartree term $Un/2$ with increasing $|k_0|$. 
The overall shape of the self-energy is the same for all interaction
strengths, only the size increases with $|U|$.

The slope of $\Im\Sg^{\Lam}(k_0)$ at $k_0 = 0$ yields the quasi particle
weight $Z_f^{\Lam}$ as
\begin{equation} \label{Zf}
 Z_f^{\Lam} = \left[ 1 - \left. 
 \partial_{k_0} \Im\Sg^{\Lam}(k_0) \right|_{k_0=0} \right]^{-1} \, .
\end{equation}
$Z_f = Z_f^{\Lam=0}$ ranges from $Z_f = 0.96$ for $U=-1.5$ to $Z_f=0.87$ 
for $U=-3$.
Although the normal self-energy is fairly small and the quasi particle
weight is only slightly suppressed for small to moderate interactions,
it has nevertheless a significant impact on the size of the pairing
gap.

Flows of the gap $\Delta^{\Lam}(k_0)$ at $k_0 = 0$ are shown in 
Fig.~15 for various choices of $U$.
\begin{figure}[tb]
\centerline{\includegraphics[width=8cm]{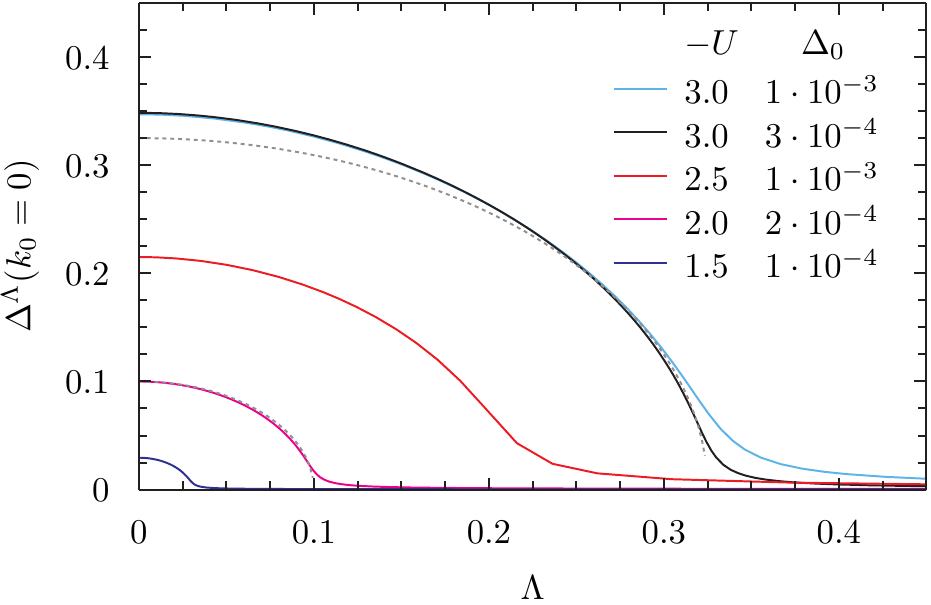}}
\caption{(Color online)
 Scale dependence of the gap $\Delta^{\Lam}(k_0)$ at $k_0 = 0$ 
 for $t'=-0.1$, $n=0.5$ and various choices of $U$ and $\Delta_0$. 
 For $U = -2$ and $-3$, the mean-field scale-dependence 
 $\sqrt{\Lam_c^2 - \Lam^2}$, with $\Lam_c$ determined from the peak 
 of $A^{\Lam}(0)$, is shown for comparison.}
\end{figure}
The small external pairing field $\Delta_0$ increases to much larger 
gaps at scales near and below the critical scale $\Lam_c$, where
$A^{\Lam}(0)$ has a peak. 
The edge of the gap flow at $\Lam_c$ becomes sharper for smaller 
$\Delta_0$.
The gap at the end of the flow assumes values close to $\Lam_c$.
The scale dependence for $\Lam < \Lam_c$ obeys approximately
\begin{equation} \label{gapflow_mf}
 \Delta^{\Lam}(0) \approx \sqrt{\Lam_c^2 - \Lam^2} \, ,
\end{equation}
with increasing accuracy for smaller values of $U$ and $\Delta_0$.
In mean-field theory this relation is exact for $\Delta_0 \to 0$, 
as one can easily see by writing down the gap equation in the 
presence of the infrared regulator Eq.~(\ref{regulator}).
For $U = -2$ the gap flow lies almost on top of the square root
function Eq.~(\ref{gapflow_mf}) for small $\Delta_0$, while for
stronger attractions deviations become visible. In particular, 
the final gap $\Delta^{\Lam=0}$ becomes clearly larger than 
$\Lam_c$.

The flow in Fig.~15 was obtained with frequency dependent 
effective boson propagators and fermion-boson vertices, and a 
frequency dependent normal self-energy and gap as described in 
Sec.~\ref{paramet}.
Comparing with results obtained by discarding the frequency
dependence of some of these quantities, one finds that only the
frequency dependence of the boson propagators and of the imaginary
part of the normal self-energy have a substantial impact on the
size of $\Delta^{\Lam}(0)$. The feedback of the other frequency
dependences on the gap at $k_0 = 0$ is small.

The critical scale and the final gap are strongly reduced compared 
to their mean-field values $\Lam_c^{\rm MF}$ and $\Delta_{\rm MF}$, 
respectively, mostly due to fluctuations {\em above} the critical 
scale.
In Fig.~16 we plot the ratio $\Delta/\Delta_{\rm MF}$ with
$\Delta = \Delta^{\Lam=0}(0)$ as a function of $U$ for $t' = -0.1$ and 
$n = 0.5$.
\begin{figure}[tb]
\centerline{\includegraphics[width=8cm]{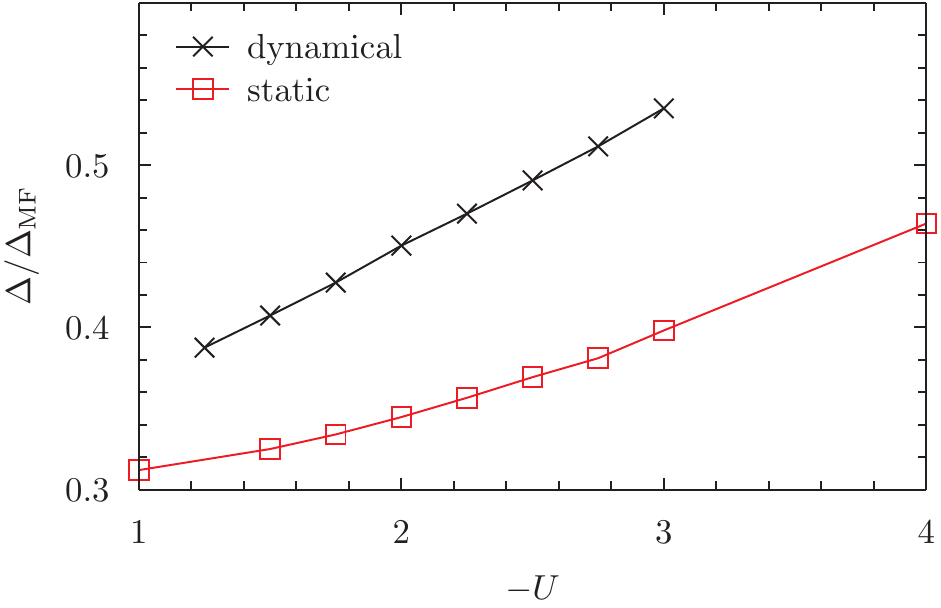}}
\caption{(Color online)
 Gap ratio $\Delta/\Delta_{\rm MF}$ as a function of $U$
 as obtained from the flow with frequency dependent (upper curve)
 and static (lower curve) vertices and self-energies.}
\end{figure}
The lower curve was obtained by a simplified static parametrization 
of the vertex and self-energy, where all frequency dependences 
where neglected.
Notably the reduction increases at weaker interactions and does not
extrapolate to one for $U \to 0$. 
This is actually the expected behavior.
In the weak coupling limit the gap $\Delta$ has the same exponential 
$U$-dependence $\Delta \propto e^{-b/|U|}$ with a (density-dependent) 
constant $b$ as in mean-field theory.
However, the prefactor of the BCS mean-field formula is reduced by 
fluctuations, as first noted for the transition temperature in 
three-dimensional superconductors by Gorkov and Melik-Barkhudarov.
\cite{gorkov61}
The reduction factor in the weak coupling limit can be computed by 
second order perturbation theory.\cite{georges91,martin92,neumayr03}
For the parameters used in Fig.~16 one finds  
$\Delta/\Delta_{\rm MF} \to 0.3$ for $U \to 0$.\cite{eberlein13}
Both curves in Fig.~16 should tend to that value, since the flow
captures the perturbative contributions. However, we cannot reach
the limit $U \to 0$ numerically. It is hard to compute the gap 
from a numerical solution of the flow equations for smaller 
interaction strengths than those shown, since $\Lam_c$ and $\Delta$ 
decrease exponentially.

For strong attractions $U$ the attractive Hubbard model can be 
mapped to a Heisenberg model in a uniform magnetic field.
\cite{micnas90} The gap ratio $\Delta/\Delta_{\rm MF}$ thereby
translates to the ratio between the staggered magnetization $m_s$
and the corresponding classical result $m_s^{\rm cl}$.
From numerical results for that ratio \cite{trivedi89} one can 
infer that the gap ratio in the strongly attractive Hubbard model 
is $0.6$ at half-filling and even larger away from half-filling.
The observed increase of $\Delta/\Delta_{\rm MF}$ with increasing 
$|U|$ is therefore consistent with the expected trend.
Similar values for $\Delta/\Delta_{\rm MF}$ but with a less
pronounced $U$-dependence have been obtained in an earlier fRG 
study with a simpler parametrization of the vertex.\cite{gersch08}

We now discuss the frequency dependence of the gap function.
In Fig.~17, $\Delta^{\Lam=0}(k_0)$ is plotted as a function of 
frequency for $U=-2$, $t'=-0.1$, and $n=1/2$.
\begin{figure}[tb]
\centerline{\includegraphics[width=8cm]{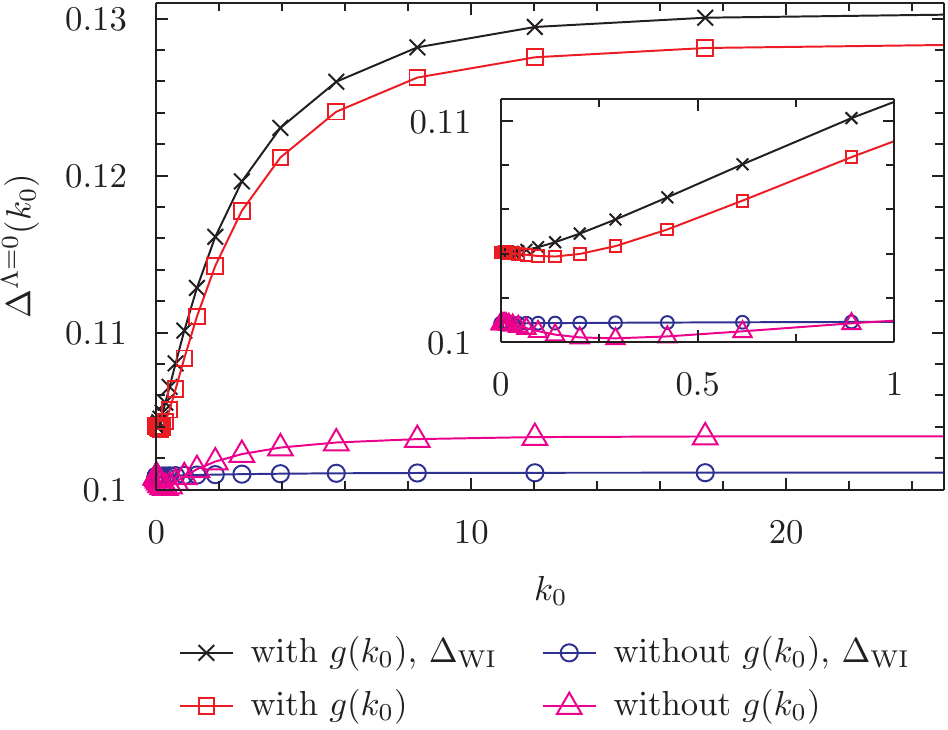}}
\caption{(Color online)
 Frequency dependence of the gap at the end of the flow
 for various approximation schemes. The inset shows the gap at
 small frequencies $k_0 \leq 1$. 
 The model parameters are $U=-2$, $t'=-0.1$, and $n=1/2$.}
\end{figure}
Results obtained by computing the gap from a projected flow 
obeying the Ward identity at $k_0 = 0$ are compared to results
where the frequency dependence of the gap is computed directly 
from the Ward identity ($\Delta_{\rm WI}$), contrasting also 
calculations with and without frequency dependent fermion-boson
vertices $g(k_0)$.
Note that the results discussed so far were all obtained by 
enforcing the Ward identity only at $k_0 = 0$.
Unlike the frequency dependence of the normal self-energy,
the frequency dependence of the gap is strongly affected by the
frequency dependent renormalization of the fermion-boson vertices.
Neglecting it leads to a very weak or almost no (for $\Delta_{\rm WI}$) 
frequency dependence.
The gap $\Delta(k_0)$ computed by projecting the flow on the Ward 
identity at $k_0 = 0$ exhibits a shallow finite-frequency 
minimum, which is probably an artifact of the approximations
associated with a (slight) violation of the Ward identity at
finite frequencies. 
$\Delta_{\rm WI}(k_0)$ has a minimum at $k_0 = 0$. A qualitatively 
similar frequency dependence of the gap is also captured by
the $T$-matrix approximation.\cite{keller99}

%%%%%%%%%%%%%%%%%%%%%%%%%%%%%%%%%%%%%%%%%%%%%%%%%%%%%%%%%%%%%%%%%%%

\subsubsection{Ward identity}
\label{wardidentity}

For real gaps $\Delta_0$ and $\Delta^{\Lam}$ the Ward identity,
Eq.~(\ref{ward1}), can be simplified to
\begin{eqnarray} \label{ward2}
 \Delta^{\Lam}(k) - \Delta_0(k) &=& 
 - \sum_{k'} \Delta_0(k') \, 
 \left[ G^{\Lam}(k') G^{\Lam}(-k') + \left(F^{\Lam}(k')\right)^2
 \right] \nonumber \\
 && \times \,
 \left[ V^{\Lam}(k,-k,-k',k') - W^{\Lam}(k,k',k',k) \right] \, .
\end{eqnarray}
Expressing $V^{\Lam}$ and $W^{\Lam}$ by the
coupling functions introduced in the channel decomposition
(Sec.~IV), the identity can be written as
\begin{equation} \label{ward3}
 \Delta^{\Lam}(k) =
 - \sum_{k'} \Delta_0(k') \, 
 \left[ G^{\Lam}(k') G^{\Lam}(-k') + \left(F^{\Lam}(k')\right)^2
 \right] \Phi_{kk'}^{\Lam}(0) + \cO(\Delta_0) \, ,
\end{equation}
for $\Delta_0 \to 0$. The first term on the right hand side is 
of order one, since $\Phi_{kk'}^{\Lam}(0) \propto \Delta_0^{-1}$ 
for small $\Delta_0$. 
With the approximate parametrization for the Hubbard model
described in Sec.~\ref{paramet}, the combination of interaction
terms on the right hand side of Eq.~(\ref{ward2}) can be 
written as
\begin{eqnarray}
 V^{\Lam}(k,-k,-k',k') - W^{\Lam}(k,k',k',k) &=&
 U + \Phi^{\Lam}(0) g^{\Lam}_{\phi}(k_0) g^{\Lam}_{\phi}(k'_0)
 \nonumber \\
 + \, \frac{1}{2} A^{\Lam}(k'-k) \left[g^{\Lam}_a(p_0)\right]^2
 &-& \frac{1}{2} \Phi^{\Lam}(k'-k) \left[g^{\Lam}_{\phi}(p_0)\right]^2
 \nonumber \\
 + \, C^{\Lam}(k'-k) \left[g^{\Lam}_c(p_0)\right]^2
 &-& 3M^{\Lam}(k'-k) \left[g^{\Lam}_m(p_0)\right]^2 \, ,
\end{eqnarray}
where $p_0 = (k_0 + k'_0)/2$.
For a small constant $\Delta_0$ and a momentum-independent
gap function $\Delta^{\Lam}(k_0)$, the Ward identity then 
assumes the form
\begin{equation} \label{ward4}
 \Delta^{\Lam}(k_0) =
 - \sum_{k'} \Delta_0 
 \left[ G^{\Lam}(k') G^{\Lam}(-k') + \left(F^{\Lam}(k')\right)^2
 \right] 
 \Phi^{\Lam}(0) g^{\Lam}_{\phi}(k_0) g^{\Lam}_{\phi}(k'_0)
 + \cO(\Delta_0) \, .
\end{equation}

The most important consequence of the Ward identity is the
divergence of the phase coupling $\Phi^{\Lam}(0)$ in the 
limit $\Delta_0 \to 0$ for $\Lam < \Lam_c$, reflecting the 
massless Goldstone boson associated with spontaneous
symmetry breaking.
The truncated flow equations do not obey the Ward identity
exactly, and $\Phi^{\Lam}(0)$ deviates from the expected
behavior $\propto \Delta_0^{-1}$ for small $\Delta_0$.
For small $U$ the deviations are tiny. 
For example, for $t'=-0.1$, $n=1/2$, and $U=-2$, the 
product $\Delta_0 \Phi^{\Lam=0}(0)$ is almost constant down 
to fairly small values of $\Delta_0$, before it increases
and finally diverges at a finite $\Delta_0$ of the order 
$10^{-5}$, which is four orders of magnitude smaller than 
$\Delta^{\Lam=0}$.
The same behavior was observed already in more pronounced
form in Ref.~\onlinecite{gersch08}.

The violation of the Ward identity can be quantified by
comparing the gap $\Delta_{\rm RG}^{\Lam}$ computed from 
its flow equation to the gap $\Delta_{\rm WI}^{\Lam}$ 
required by the Ward identity. 
The latter is computed from Eq.~(\ref{ward2}) by inserting
the coupling functions as determined from the flow on the 
right hand side. In Fig.~18 we plot the difference 
$\Delta_{\rm RG}^{\Lam} - \Delta_{\rm WI}^{\Lam}$, divided 
by $\Delta_{\rm RG}^{\Lam} U^4$, as a function of the scale 
in units of $\Lam_c$.
\begin{figure}[tb]
\centerline{\includegraphics[width=8cm]{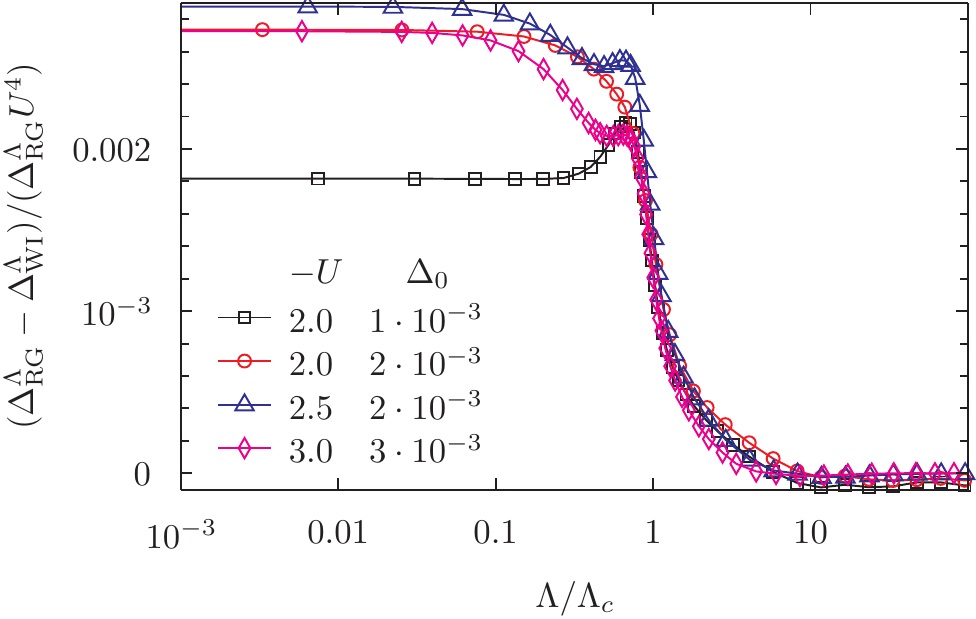}}
\caption{(Color online)
 Violation of the Ward identity as a function of
 the scale $\Lam$ for $t'=-0.1$, $n=1/2$ and various values 
 of $U$ and $\Delta_0$. 
 The gap $\Delta_{\rm RG}^{\Lam}$ determined from the flow
 equation is compared to the gap $\Delta_{\rm WI}^{\Lam}$
 determined from the Ward identity.}
\end{figure}
One can see that the violation builds up gradually at 
scales around $\Lam_c$. The normalized difference
$(\Delta_{\rm RG}^{\Lam} - \Delta_{\rm WI}^{\Lam})/
 \Delta_{\rm RG}^{\Lam}$
increases rapidly from $U=-2$ to $U=-3$.
For $\Lam > \Lam_c$ it is roughly proportional to $U^4$.
For $\Lam < \Lam_c$ a pronounced $\Delta_0$-dependence
appears. For much smaller values of $\Delta_0$ than those
shown, $\Delta_{\rm RG}^{\Lam} - \Delta_{\rm WI}^{\Lam}$ can turn 
negative, which is related to an artificial divergence of 
$\Phi^{\Lam}$ at a finite $\Delta_0$.
We observed similar $U$-dependences also for other hopping 
parameters and densities.
On general grounds one would expect a violation of the
Ward identity of order $U^3$ at weak coupling, even if
the one-loop flow was carried out without additional
approximations.\cite{katanin04,eberlein13}
The above results suggest that the violation sets in only 
at order $U^4$, or contributions of order $U^3$ have very 
small prefactors.

In Fig.~19, 
$(\Delta_{\rm RG}^{\Lam} - \Delta_{\rm WI}^{\Lam})/
 \Delta_{\rm RG}^{\Lam}$
is plotted as a function of $\Lam$ for a fixed set of
parameters, to compare the performance of different 
approximations.
\begin{figure}[tb]
\centerline{\includegraphics[width=8cm]{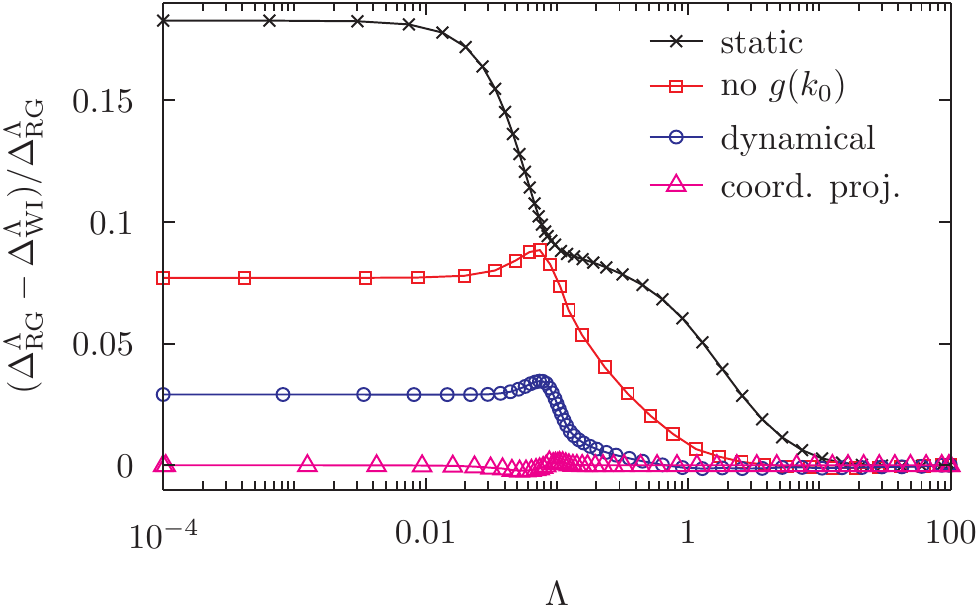}}
\caption{(Color online)
 Violation of the Ward identity as a function of $\Lam$
 for $U=-2$, $t'=-0.1$, $n=1/2$, and $\Delta_0 = 10^{-3}$. 
 Results from the flow equations with static and two distinct 
 (with and without frequency dependent fermion-boson vertices) 
 dynamical parametrizations of the vertex are compared to the 
 result from the Ward identity projected flow.}
\end{figure}
The graph labeled ''dynamical'' was obtained by using the
frequency dependent parametrization of the vertex and the
self-energy as described in Sec.~\ref{paramet}. 
The graph ''no $g(k_0)$'' was computed with constant (unity) 
fermion-boson vertices, and the graph ''static'' by discarding
all frequency dependences.
The lowest curve labeled ''coord.\ proj.'' was computed with 
the dynamical parametrization and the Ward identity enforced
by ''coordinate projection'' (as described below).
The latter obeys the Ward identity by construction, up to
small discretization errors.
Taking the frequency dependences into account obviously
reduces the violation of the Ward identity significantly.

Even for the most accurate parametrization of the vertex, 
the Ward identity is not fulfilled by the truncated flow, 
as generally expected.\cite{katanin04}
A detailed discussion of this problem in the case of
superfluid order is provided in Ref.~\onlinecite{eberlein13}.
The deviations are small for weak interactions but increase
rapidly with $|U|$.
Violating the Ward identity spoils the singular infrared
behavior of the coupling functions associated with the 
massless Goldstone boson for $\Delta_0 \to 0$. 
Even worse, it leads to artificial singularities which
prevent one from carrying out the flow down to $\Lam \to 0$
and $\Delta_0 \to 0$.
In the results presented in the preceding sections we have
therefore enforced the Ward identity by using a coordinate
projection procedure, devised for the numerical solution of
systems of ordinary differential equations with constraints.
\cite{ascher94}
The flowing quantities are thereby projected on the manifold 
spanned by the constraint (Ward identity) in a way that the 
projected solution stays as close to the solution of the 
flow equations as possible, while deviations from the 
constraint are damped exponentially.
In practice, we have enforced the Ward identity only at zero 
frequency ($k_0 = 0$), to reduce the numerical effort.
\cite{eberlein13}
This has little effect on absolute values of results, but
leads to the slightly artificial frequency dependence of the 
gap at low frequencies discussed in Sec.~\ref{selfgap}.

%%%%%%%%%%%%%%%%%%%%%%%%%%%%%%%%%%%%%%%%%%%%%%%%%%%%%%%%%%%%

\subsubsection{$\Delta_0$-flow and singularities}
 \label{delta0flow}

We finally take a closer look at the singularities of the 
vertex in the limit $\Delta_0 \to 0$.
In particular, we complement the numerical results for the
Hubbard model by qualitative analytical estimates which 
are generally valid for fully gapped singlet superfluids.

To this end, we assume that the fermionic cutoff has already
been removed ($\Lam \to 0$), and we analyze the flow as a
function of a decreasing pairing field $\Delta_0$.
In Fig.~20 we show the flow of $\Delta(0) = \Delta^{\Lam=0}(0)$, 
$\Phi(0) = \Phi^{\Lam=0}(0)$ and $A(0) = A^{\Lam=0}(0)$ as a 
function of $\Delta_0$, with an initial value $\Delta_0 = 0.005$. 
Results obtained for some fixed smaller values of $\Delta_0$ 
are shown for comparison.
\begin{figure}[tb]
\centerline{\includegraphics[width=8cm]{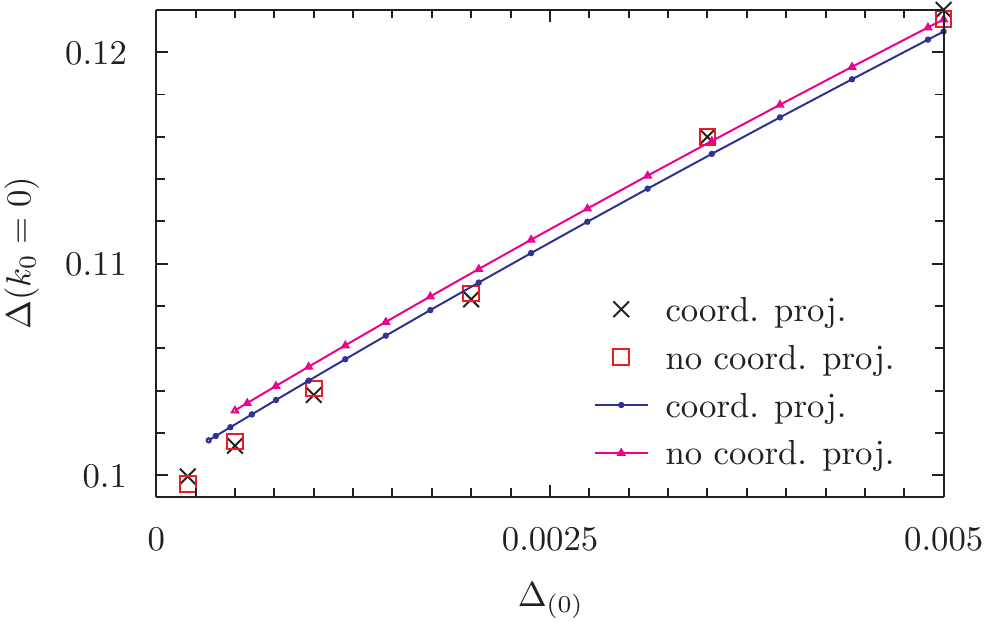}}
\centerline{\includegraphics[width=8cm]{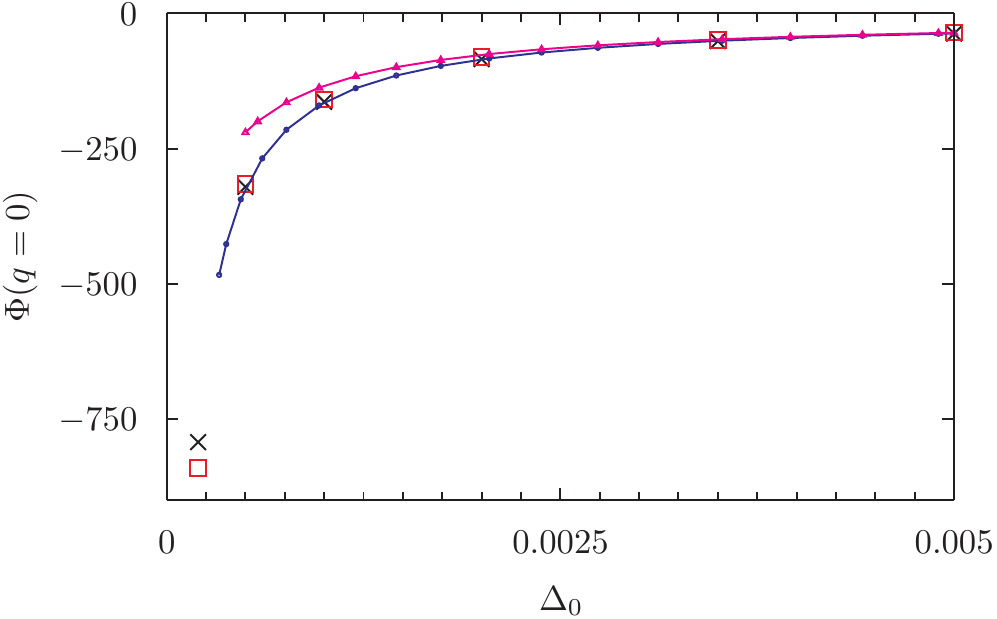}}
\centerline{\includegraphics[width=8cm]{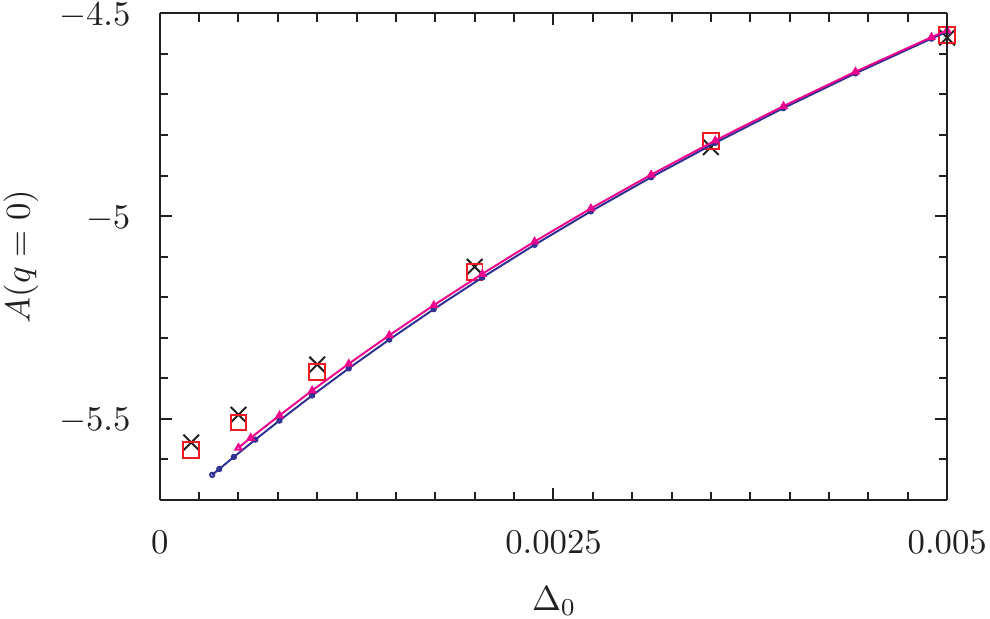}}
\caption{(Color online)
 $\Delta_0$-flows of $\Delta(0)$, $\Phi(0)$ and
 $A(0)$ for an initial value $\Delta_0 = 0.005$. 
 Results obtained for fixed smaller values of $\Delta_0$
 are shown for comparison (symbols).
 The model parameters are $t'=-0.1$, $n=0.5$, $U = -2$.}
\end{figure}
The numerical computation of the $\Lam$-flow becomes 
increasingly difficult at smaller $\Delta_0$.
Furthermore, there are systematic deviations between the 
results obtained from the $\Delta_0$-flow and those 
computed at fixed $\Delta_0$.
These may be related to divergencies in box diagrams
for $\Delta_0 \to 0$, which can and must be treated by
a flow starting at an initially finite $\Delta_0$, as we 
discuss in the following.

In a fully gapped superfluid, the fermionic propagator 
is regularized by the pairing gap. 
However, the interaction vertex develops a singularity
associated with the emergence of a Goldstone boson.
In particular, the phase coupling function has the singular
form
\begin{equation} \label{phising}
 \Phi(q) \propto 
 - \frac{1}{\Delta_0 + a q_0^2 + b \bq^2} \; ,
\end{equation}
for small $q=(q_0,\bq)$, where $a$ and $b$ are positive 
constants.
This singularity is dictated by the Ward identity.
Related singularities occur also for the imaginary parts
of the pairing and anomalous (3+1)-coupling functions
$P''(q)$ and $X''(q)$, respectively, but their impact is 
reduced by a numerator proportional to $q_0$.
The divergence of $\Phi(q)$ for $q \to 0$, $\Delta_0 \to 0$ 
is integrable in (2+1) dimensions. 
Hence, self-energy and vertex correction (Fig.~3) 
contributions involving integrals over $\Phi(q)$ remain 
finite for $\Delta_0 \to 0$.
However, box diagrams (Fig.~4) yield contributions involving
an integral over products of two phase coupling functions,
\begin{equation}
 \sum_p \partial_{\Delta_0} \left[
 G_{s_1s_2}(p-q/2) G_{s_3s_4}(p+q/2) \right] \,
 \Phi(p-k) \Phi(k'-p) \, ,
\end{equation}
where $G$ is the propagator and $\Phi$ the phase coupling
function for $\Lam=0$.
The $\Delta_0$-derivative of the propagators is finite for
$\Delta_0 \to 0$, but for $k=k'$ the singularities of the 
phase coupling functions coalesce and the integral diverges 
as $\Delta_0^{-1/2}$.
It is therefore not possible to set $\Delta_0$ to zero
before the fermionic cutoff has been removed.
The $\Delta_0$-flow is however well defined and integrable.

Hence, the singularities associated with the Goldstone mode
do not lead to divergencies in other channels.
In this respect the one-loop flow analyzed in this work is
qualitatively similar to the RPA. The fluctuation effects
beyond RPA yield only finite renormalizations.
On the other hand, it is known from the theory of interacting
bosons that the phase mode does lead to a singular renormalization
of the amplitude mode.\cite{pistolesi04}
In a renormalization group theory of fermionic superfluids with
auxiliary boson fields representing the order parameter 
fluctuations, this effect appears already at one-loop level.
\cite{strack08}
The singular contributions involve scale derivatives acting
on the boson propagators.
In the purely fermionic renormalization group (without 
auxiliary bosons), analogous singular contributions appear
only at the two-loop level.\cite{eberlein13}

%%%%%%%%%%%%%%%%%%%%%%%%%%%%%%%%%%%%%%%%%%%%%%%%%%%%%%%%%%%%%%%%%%%

\section{Conclusion}

We have analyzed ground state properties of a spin-singlet 
superfluid including fluctuations on all scales via a fermionic 
functional renormalization group flow in a formulation that 
allows for symmetry breaking.
The flow equations were truncated in a one-loop approximation 
with self-energy feedback.
Spin rotation invariance and discrete symmetries were fully 
exploited to simplify the structure of the Nambu two-particle 
vertex.
To parametrize the singular momentum and frequency dependences
of the effective interactions, the Nambu vertex was decomposed 
in charge, magnetic, and various normal and anomalous pairing 
channels, which are all mutually coupled in the flow.
We have shown that the channel decomposed one-loop flow
equations are equivalent to the RPA for the vertex and to 
mean-field theory for the gap function, if only direct 
Nambu particle-hole contributions are taken into account.
\cite{fn3} 
The crossed particle-hole and the particle-particle (in Nambu 
representation) contributions to the complete one-loop flow 
thus capture fluctuations beyond mean-field theory and RPA.

We have evaluated the flow equations for the two-dimensional 
attractive Hubbard model as a prototype of an interacting Fermi
system with a spin-singlet superfluid ground state. 
The dominance of $s$-wave terms in the effective interactions
in that model allows for a relatively simple parametrization.
The global $U(1)$ Ward identity relating the vertex to the gap 
function is violated by the one-loop truncation. The deviations
are very small for a weak attraction, but increase rapidly for 
stronger interactions. To maintain the singularity structure 
associated with the Goldstone boson, the flow was therefore 
projected on the Ward identity, analogously to evaluating a 
differential flow in the presence of a constraint.
We have computed the effective interactions in the charge,
magnetic, and pairing channels, including anomalous 
(3+1)-interactions describing pair annihilation (or creation) 
combined with a one-particle scattering process. 
Unprecedented comprehensive results on the momentum and 
(imaginary) frequency dependences of the effective interactions 
were obtained and discussed.
The singularities in the pairing channels generated by the 
one-loop flow are qualitatively similar to the RPA, and are 
to a large extent fixed by the Ward identity. 
The effective magnetic interaction develops a low-frequency 
small-momentum peak which is a pure fluctuation effect.
There are also significant quantitative fluctuation effects
which are captured by the one-loop flow. In particular, the
gap is strongly reduced compared to the mean-field value,
with a stronger reduction at weaker interactions, as expected
from perturbative and numerical results.
The expected divergence of the superfluid amplitude mode in the
low-energy limit is not captured by the one-loop truncation.
This effect appears only at the two-loop level in the fermionic 
renormalization group flow.\cite{eberlein13}

Besides the channel decomposition of the vertex for a system 
exhibiting spontaneous symmetry breaking, there are two other 
noteworthy technical upshots of our work, which may be picked
up in future calculations.
First, we have found that an accurate discretization of both 
momentum and frequency dependences is computationally feasible
\cite{fn4} and has several advantages compared to the usual 
strategy of an ansatz with a small number of scale-dependent 
coefficients.
In particular, one avoids problems with momentum or frequency 
derivatives which are necessary to extract the flow of such 
coefficients.
Second, we have shown that a symmetry breaking field can be 
used as a convenient flow parameter, which regularizes the flow 
at the critical scale and allows for a controlled treatment 
of infrared divergences associated with the Goldstone boson.

%%%%%%%%%%%%%%%%%%%%%%%%%%%%%%%%%%%%%%%%%%%%%%%%%%%%%%%%%%%%%%%%%%%%

\begin{acknowledgments}
We are grateful to J.~Bauer, K.-U.~Giering, N.~Hasselmann, 
T.~Holder, C.~Husemann, A.~Katanin, B.~Obert, and M.~Salmhofer 
for valuable discussions. 
\end{acknowledgments}

%%%%%%%%%%%%%%%%%%%%%%%%%%%%%%%%%%%%%%%%%%%%%%%%%%%%%%%%%%%%%%%%%%%%%

\begin{appendix}

\section{Pauli matrix basis}

It is often convenient to represent the Nambu vertex in a basis spanned 
by tensor products of Pauli matrices and the unit matrix.
The Pauli matrices $\tau^{(1)}$, $\tau^{(2)}$, $\tau^{(3)}$ and the unit 
matrix $\tau^{(0)}$ form a basis in the vector space of complex 
$2 \times 2$ matrices. 
The tensor products $\tau^{(j)} \otimes \tau^{(j')}$ form a basis in 
the space of complex $4 \times 4$ matrices.
The components of the Nambu vertex in this basis are obtained as
\begin{equation}
 \tilde\Gam^{(4)\Lam}_{jj'}(k_1,k_2,k_3,k_4) = 
 \frac{1}{2} \sum_{s_i} \tau^{(j)}_{s_4s_1} \tau^{(j')}_{s_3s_2} \,
 \Gam^{(4)\Lam}_{s_1s_2s_3s_4}(k_1,k_2,k_3,k_4) \; .
\end{equation}
The inverse basis transformation is given by
\begin{equation}
 \Gam^{(4)\Lam}_{s_1s_2s_3s_4}(k_1,k_2,k_3,k_4) =
 \frac{1}{2} \sum_{j,j'} 
 \tau^{(j)}_{s_1s_4} \tau^{(j')}_{s_2s_3} \,
 \tilde\Gam^{(4)\Lam}_{jj'}(k_1,k_2,k_3,k_4) \; .
\end{equation}
The matrix formed by the components $\tilde\Gam^{(4)\Lam}_{jj'}$ is
denoted as $\tilde\bGam^{(4)\Lam}$. The tilde is used to distinguish 
this and other matrices represented in the Pauli basis from matrices 
in the Nambu index basis defined in Eq.~(\ref{vertexmatrixdef}).

The flow equations for the coupling functions parametrizing the
channel decomposed Nambu vertex can be derived most conveniently
in the Pauli matrix basis. 
Since the complete set of coupling functions is contained in the
particle-hole contribution to the vertex, their flow is determined
by the flow equation Eq.~(\ref{flowVPH}). Transformed to the Pauli
matrix basis, the equation reads
\begin{equation}
 \frac{d}{d\Lam} \tilde V^{\rm PH,\Lam}_{jj'}(k,k';q) =
 \sum_p \sum_{l,l'} \tilde\Gam^{(4)\Lam}_{jl}(k,p;q) \,
 \partial_{\Lam} \tilde L^{\Lam}_{ll'}(p;q) \,
 \tilde\Gam^{(4)\Lam}_{l'j'}(p,k';q) \; ,
\end{equation}
where
\begin{equation}
 \tilde L^{\Lam}_{jj'}(p;q) = \frac{1}{2} \sum_{s_i}
 \tau^{(j)}_{s_4s_1} \tau^{(j')}_{s_3s_2} \,
 G^{\Lam}_{s_2s_4}(p-\q2) G^{\Lam}_{s_1s_3}(p+\q2) \; .
\end{equation}
The decomposition Eq.~(\ref{vertexdecomp}) of the Nambu vertex 
can be written in the Pauli matrix basis with momentum variables
$k_{1,4} = k \pm q/2$ and $k_{2,3} = k' \mp q/2$ as
\begin{eqnarray}
 \tilde\Gam^{(4)\Lam}_{jj'}(k,k';q) &=& 
 \tilde U_{jj'}(k,k';q) + 
 \tilde V^{\rm PH,\Lam}_{jj'}(k,k';q) 
 \nonumber \\ &-&
 \tilde V^{\rm PH',\Lam}_{jj'}
 \left( \textstyle{\frac{k+k'-q}{2},\frac{k+k'+q}{2};k'-k} \right)
 \nonumber \\ &+&
 \tilde V^{\rm PP,\Lam}_{jj'}
 \left( \textstyle{\frac{k-k'+q}{2},\frac{k-k'-q}{2};k+k'} \right)
 \; .
\end{eqnarray}
Note that $\tilde V^{\rm PH',\Lam}_{jj'}$ is defined by transforming
$V^{\rm PH,\Lam}_{s_2s_1s_3s_4}$ with the first two Nambu indices 
exchanged to the Pauli matrix basis.
The functions $\tilde L^{\Lam}_{jj'}(p;q)$ are given by products of
normal and anomalous propagators,
\begin{eqnarray}
 L^{\Lam}_{00}(p;q) &=& 
 \Re[G^{\Lam}(p_-) G^{\Lam}(p_+)] + F^{\Lam}(p_-) F^{\Lam}(p_+) \; ,
 \nonumber \\
 L^{\Lam}_{01}(p;q) &=& 
 iF^{\Lam}(p_-) \Im G^{\Lam}(p_+) + i\Im G^{\Lam}(p_-) F^{\Lam}(p_+)
 = L^{\Lam}_{10}(p;q) \; ,
 \nonumber \\
 L^{\Lam}_{02}(p;q) &=& 
 i\Re G^{\Lam}(p_-) F^{\Lam}(p_+) - iF^{\Lam}(p_-) \Re G^{\Lam}(p_+)
 = - L^{\Lam}_{20}(p;q) \; ,
 \nonumber \\
 L^{\Lam}_{03}(p;q) &=& 
 i\Im[G^{\Lam}(p_-) G^{\Lam}(p_+)] = L^{\Lam}_{30}(p;q) \; ,
 \nonumber \\
 L^{\Lam}_{11}(p;q) &=& 
 - \Re [G^{\Lam}(p_-) G^{\Lam *}(p_+)] 
 + F^{\Lam}(p_-) F^{\Lam}(p_+) \; ,
 \nonumber \\
 L^{\Lam}_{22}(p;q) &=& 
 - \Re [G^{\Lam}(p_-) G^{\Lam *}(p_+)] 
 - F^{\Lam}(p_-) F^{\Lam}(p_+) \; ,
 \nonumber \\
 L^{\Lam}_{33}(p;q) &=& 
 \Re[G^{\Lam}(p_-) G^{\Lam}(p_+)] - F^{\Lam}(p_-) F^{\Lam}(p_+) \; ,
 \nonumber \\
 L^{\Lam}_{12}(p;q) &=& 
 \Im [G^{\Lam}(p_-) G^{\Lam *}(p_+)] = 
 - L^{\Lam}_{21}(p;q) \; ,
 \nonumber \\
 L^{\Lam}_{13}(p;q) &=& 
 \Re G^{\Lam}(p_-) F^{\Lam}(p_+) + F^{\Lam}(p_-) \Re G^{\Lam}(p_+) 
 = L^{\Lam}_{31}(p;q) \; ,
 \nonumber \\
 L^{\Lam}_{23}(p;q) &=& 
 \Im G^{\Lam}(p_-) F^{\Lam}(p_+) - F^{\Lam}(p_-) \Im G^{\Lam}(p_+) 
 = - L^{\Lam}_{32}(p;q) \; ,
\end{eqnarray}
where $p_+ = p + q/2$, $p_- = p - q/2$.

The matrices representing the Nambu vertex in our approximation for
the Hubbard model are not all full, that is, several matrix elements
vanish. More generally, for coupling functions with a factorized 
momentum dependence and even parity form factors the (direct)
particle-hole contribution to the vertex has the form
\begin{equation}
 \tilde\bV^{\rm PH,\Lam}(k,k';q) =
 \left( \begin{array}{cccc}
 2 M^{\Lam}_{kk'}(q) & 0 & 0 & 0 \\
 0 & A^{\Lam}_{kk'}(q) & P''^{\Lam}_{kk'}(q) & 2X'^{\Lam}_{kk'}(q) \\
 0 & - P''^{\Lam}_{kk'}(q) & \Phi^{\Lam}_{kk'}(q) &  - 2X''^{\Lam}_{kk'}(q) \\
 0 & 2X'^{\Lam}_{kk'}(q) & 2X''^{\Lam}_{kk'}(q) & 2 C^{\Lam}_{kk'}(q)
 \end{array} \right) \; ,
\end{equation}
and the particle-particle contribution $\tilde\bV^{\rm PP,\Lam}$ has
only diagonal elements given by the magnetic coupling function
$M^{\Lam}_{kk'}(q)$. However, the matrix elements of the crossed 
particle-hole contribution $\tilde\bV^{\rm PH',\Lam}(k,k';q)$ are
all non-zero and generally given by linear combinations of several
coupling functions.

\end{appendix}

%%%%%%%%%%%%%%%%%%%%%%%%%%%%%%%%%%%%%%%%%%%%%%%%%%%%%%%%%%%%%%%%%%%%%


\begin{thebibliography}{99}
\bibitem{fn2} In the matrix $\bGam^{(4)}$ presented in 
 Ref.~\onlinecite{eberlein10}, Eq.~(3.18), the order of the first
 two momenta of the matrix elements $\Gam^{(4)}_{-+++}$ and
 $\Gam^{(4)}_{---+}$ was erroneously reversed.
\bibitem{fn3} 
 Recall that the Cooper (particle-particle) contributions are 
 transformed to particle-hole terms in Nambu representation.
\bibitem{fn4}
 For flows in the symmetric regime a treatment of momentum and 
 frequency dependences by discretization has been used already 
 by Giering and Salmhofer, see Ref.~\onlinecite{giering12}.
\end{thebibliography}
\end{document}